\documentclass[fleqn]{2023SCGE}
\usepackage{aas_macros}
\usepackage[square,sort,comma,numbers]{natbib}
\usepackage{amssymb,amsmath}
\usepackage{xspace}
\usepackage{amsmath}
\usepackage{longtable}
\usepackage{makecell}
\usepackage[scheme=plain]{ctex}

\newcommand\msunh{h^{-1}\mathrm{M}_\odot}

\newcommand{\mpch}{h^{-1}{\rm Mpc}}
\newcommand{\kpch}{h^{-1}{\rm kpc}}

\newcommand{\msun}{\rm M_\odot}
\newcommand{\hbt}{\textsc{hbt+}\xspace}
\newcommand{\fof}{\textsc{FoF}\xspace}

\newcommand{\lgadget}{\textsc{LGadget}\xspace}
\newcommand{\gadget}{\textsc{Gadget}\xspace}
\newcommand{\subfind}{\textsc{SubFind}\xspace}

\newcommand{\gaea}{\textsc{gaea}\xspace}
\newcommand{\blic}{\textsc{Blic}\xspace}
\newcommand{\starduster}{\textsc{StarDuster}\xspace}

\begin{document}
\ensubject{subject}
\ArticleType{Article}
\SpecialTopic{SPECIAL TOPIC: }
\Year{2025}
\Month{January}
\Vol{66}
\No{1}
\DOI{??}
\ArtNo{000000}
\ReceiveDate{Oct 31, 2025}
\AcceptDate{Nov 4, 2025}

\title{A semi-analytical mock galaxy catalog for the CSST extragalactic surveys from the Jiutian simulations}{A semi-analytical mock galaxy catalog for the CSST extragalactic surveys from the Jiutian simulations}

\author[1,2]{Zhenlin Tan}{}
\author[3]{Lizhi Xie}{}
\author[1,2]{Jiaxin Han}{{jiaxin.han@sjtu.edu.cn}}
\author[4]{Yisheng Qiu}{}
\author[5, 6]{Fabio Fontanot}{}
\author[5]{Gabriella De Lucia}{}
\author[7]{\\Qi Guo}{}
\author[1,2]{Qingyang Li}{}
\author[1,2]{Jiale Zhou}{}
\author[1,2]{Wenkang Jiang}{}
\author[8,9,10]{Xin Wang}{}
\author[1,2]{Feihong He}{}
\author[9,11,12]{\\Chichuan Jin}{}
\author[1,2]{Yipeng Jing}{}
\author[7]{Ming Li}{}
\author[13,14,15]{Xiaodong Li}{}
\author[16]{Wenxiang Pei}{}
\author[1,2]{\\Wenting Wang}{}
\author[1,2]{Xiaohu Yang}{}
\author[1,2]{Yu Yu}{}


\AuthorMark{Tan Z., et al.}

\AuthorCitation{Tan Z., et al.}

\address[1]{Department of Astronomy, School of Physics and Astronomy, Shanghai Jiao Tong University, Shanghai, 200240, China}
\address[2]{State Key Laboratory of Dark Matter Physics, Key Laboratory for Particle Astrophysics and Cosmology (MoE), \\\& Shanghai Key Laboratory for Particle Physics and Cosmology, Shanghai Jiao Tong University, Shanghai, 200240, China}
\address[3]{Astrophysics Center, Tianjin Normal University, Tianjin, 300387, China}
\address[4]{Research Center for Astronomical Computing, Zhejiang Laboratory, Hangzhou 311121, China}
\address[5]{INAF - Astronomical Observatory of Trieste, via G.B. Tiepolo 11, I-34143 Trieste, Italy}
\address[6]{IFPU - Institute for Fundamental Physics of the Universe, via Beirut 2, 34151, Trieste, Italy}
\address[7]{Key Laboratory for Computational Astrophysics, National Astronomical Observatories, Chinese Academy of Sciences, Beijing 100101, China}
\address[8]{School of Astronomy and Space Science, University of Chinese Academy of Sciences (UCAS), Beijing 100049, China}
\address[9]{National Astronomical Observatories, Chinese Academy of Sciences, Beijing 100101, China}
\address[10]{Institute for Frontiers in Astronomy and Astrophysics, Beijing Normal University, Beijing 102206, China}
\address[11]{School of Astronomy and Space Science, University of Chinese Academy of Sciences, Beijing 100049, China.}
\address[12]{Institute for Frontier in Astronomy and Astrophysics, Beĳing Normal University, Beijing 102206, China.}
\address[13]{Peng Cheng Laboratory, Shenzhen, Guangdong 518066, China}
\address[14]{School of Physics and Astronomy, Sun Yat-Sen University, Zhuhai 519082, China}
\address[15]{CSST Science Center for the Guangdong–Hong Kong–Macau Greater Bay Area, SYSU, Zhuhai 519082, China}
\address[16]{Shanghai Key Lab for Astrophysics, Shanghai Normal University, Shanghai 200234, People’s Republic of China}

\abstract{
We introduce a mock galaxy catalog built for the CSST extragalactic surveys using the primary runs of the Jiutian $N$-body simulation suites. The catalogs are built by coupling the GAlaxy Evolution and Assembly (\gaea) semi-analytical model of galaxy formation with merger trees extracted from the simulations using the Hierarchical Bound-Tracing (\hbt) algorithm. The spectral energy distributions (SEDs) and broadband magnitudes are computed using the neural-network-based stellar population synthesizer \starduster, which is trained on radiative transfer simulations to account for detailed galaxy geometry in modeling dust obscuration. Galaxy light-cones up to $z=5$ are subsequently generated with the \blic light-cone builder which interpolates the properties of galaxies over time using an optimized interpolation scheme. The resulting catalogs exhibit good convergence in many statistical properties of the galaxy population produced from two different resolution simulations. The catalogs reproduce a number of observed galaxy properties across a range of galaxy mass and redshift, including the stellar mass functions, the luminosity function, gas mass fraction, galaxy size-mass relation and galaxy clustering. We also present the photometric and redshift distributions of galaxies expected to be observed in the CSST surveys.
}



\maketitle

\begin{multicols}{2}
\section{Introduction} \label{sec:intro}

Over the past few decades, large-scale galaxy surveys such as the Two-Degree Field Galaxy Redshift Survey~\citep[2dFGRS;][]{2001MNRAS.328.1039C}, the Sloan Digital Sky Survey~\citep[SDSS;][]{2000AJ....120.1579Y}, and the Dark Energy Spectroscopic Instrument~\citep[DESI;][]{2016arXiv161100036D} have extensively mapped the distribution of galaxies across vast expanses of our universe. These surveys have played a crucial role in shaping the current standard cosmological model and significantly advancing our understanding of galaxy formation and evolution.
At the same time, cosmological simulations such as the Millennium simulation~\citep{2005Natur.435..629S}, the EAGLE simulation~\citep{2015MNRAS.446..521S,2015MNRAS.450.1937C}, and the Illustris-TNG simulation series~\citep{Marinacci2018,Naiman2018,Nelson2018,Pillepich2018,Springel2018,Nelson2019} provide precise predictions on the matter distributions in the universe, assisting our interpretations of the observed Universe. 

To compare simulations against observations, it is crucial to account for various selection effects and observational systematics when deriving observables from simulations. A straightforward approach to accomplish this task is to forward model the observational effects in mock galaxy catalogs built from simulations.
These catalogs can then be directly compared against observational data, allowing a better understanding of the modelling of key physical mechanisms included in the simulation~\citep[e.g.][]{Diaferio1999MNRAS, Carretero2015MNRAS, Smith2017MNRAS, Wechsler2022ApJ}. Besides, mock catalogs can also serve as essential tools for evaluating and optimizing observational strategies at the survey planning stage, by quantifying observational systematics and predicting their influences on the final statistics of interest~\citep[e.g.][]{Cole1998MNRAS, Obreschkow2009ApJ, Cai2009MNRAS, Sasha2021MNRAS, Gu2024MNRAS, Shi2025SCPMA..6849511S}.

In this work, we present a mock galaxy catalog for the China Space Survey Telescope (CSST) extragalactic surveys. The CSST is a 2-meter space telescope planned to be in the same orbit as the China Manned Space Station~\cite{CSSTintro}. The telescope is designed to have a wide field of view of at least $1.7\ \mathrm{deg}^2$ and an angular resolution of at least $0.13$ arc-seconds. It will be equipped with multiple backend instruments serving different observational tasks, including a survey module capable of performing both photometric and spectroscopic observations simultaneously. With the survey module, the CSST optical survey plans to observe $\sim 17500~\mathrm{deg}^2$ of the sky in about ten years\cite{2011SSPMA..41.1441Z, 2018MNRAS.480.2178C, 2019ApJ...883..203G, ZhanHu2021...CSB, Gong2025SCPMA..6880402G}, using seven photometric imaging bands (see Figure~\ref{fig:filters}) and three spectroscopic bands covering a wavelength range of 255-1000 nm.

The construction of mock catalogs is an essential task for the preparation of the CSST surveys and for future cosmological analysis of the CSST data. 
In order to compare with the vast survey footprint and the deep redshift coverage, large-scale simulations able to track matter distributions over the Large-Scale Structure are imperative. $N$-body simulations are still the most effective approach due to their relatively low computational costs compared with hydrodynamical simulations. The next crucial step requires the generation of mock galaxy catalogs and synthetic photometry, to mimic the observational data selection and analysis.

In this work, we have developed an end-to-end pipeline for generating mock catalogs from $N$-body simulations, combining and developing a few state-of-the-art post-processing tools. The pipeline integrates the \gaea (GAlaxy Evolution and Assembly)~\citep{xie2020} semi-analytical model of galaxy formation with outputs from the \hbt (Hierarchical Bound-Tracing)~\citep{hbtplus} subhalo finder and merger tree builder. The spectral energy distributions (SEDs) of model galaxies are calculated with the fast and flexible SED modeler \starduster ~\citep{2022ApJ...930...66Q}. A generic light-cone builder, \blic, has been developed to produce continuous galaxy light-cones from the galaxy catalogs at discrete snapshots. Applying this pipeline to the Jiutian simulations~\cite{Han2025SCPMA..6809511H}, we have generated full-sky mock light-cones for the CSST survey. The resulting catalogs can reproduce many observational properties of galaxies over a range of mass and redshift, including the stellar mass and luminosity functions, gas mass fraction, galaxy size-mass relation, and galaxy clustering. The catalog is also used to forecast some basic properties of the CSST galaxies. 

The purpose of this paper is to introduce the pipeline development, to present the properties of the resulting catalog, and to serve as a documentation for using the catalog.
The layout of this paper is as follows. Section~\ref{sec:simulations} introduces the Jiutian simulations and the halo/subhalo catalogs. Section~\ref{sec:pipe} introduces the postprocessing pipeline for building the mocks, including the semi-analytical model in Section~\ref{sec:sam}, the SED generator in Section~\ref{sec:sed}, and the lightcone builder in Section~\ref{sec:lc}. Section~\ref{sec:results} presents the results of galaxy properties from our simulations, while Section~\ref{sec:forecasts} provides forecasts for CSST observations. Finally, we summarize our results and present conclusions in Section~\ref{sec:conclu}. The comparisons of light-cone interpolation schemes and descriptions of the released data, are contained in Supplementary materials.

\section{The Jiutian Simulations and Subhalo Catalogs}
\label{sec:simulations}

The Jiutian simulation suite comprises a series of $N$-body simulations specifically designed to meet the science requirements for the CSST galaxy surveys~\citep{Han2025SCPMA..6809511H}. It consists of four subsets of simulations, including the primary runs targeting the concordance cosmology, the extension runs with various non-standard cosmologies, the emulator runs covering a wide cosmological parameter space~\citep{Chen2025SCPMA..6889512C}, and the constrained runs aiming to reproduce the actual observed Large-Scale Structure~\citep{Resim2}.

Our mock catalog construction utilizes two primary simulations from the Jiutian suite (Table~\ref{tab:primary}), both employing $6144^3$ particles but differing in box sizes.\footnote{A third simulation in the primary run with a boxsize of $300\mpch$ is still being processed and thus not presented in this work.} These simulations adopt the Planck-2018 $\Lambda$CDM cosmology (from the final column of Table~2 in \citet{Planck2018}) with parameters $\Omega_m=0.3111$, $\Omega_\Lambda=0.6889$, $\Omega_b=0.049$, $n_s=0.9665$, $\sigma_8=0.8102$, and $h=0.6766$. The Jiutian-1G simulation is run with {\lgadget}-3 \citep{Angulo12}, while Jiutian-2G utilizes the newly released \gadget-4 code~\citep{Gadget4}. To generate the initial conditions, the linear power spectrum is computed with the \textsc{CAMB} code\footnote{\url{https://camb.info}} at an initial redshift of $z_{\rm ini}=127$. The Zel'dovich approximation is adopted to establish the displacement fields, deriving position and velocity perturbations from glass-like pre-initial conditions. Similar to the Millennium II simulation~\citep{MillenniumII}, these simulations output 9 snapshots from $z=127$ to $z=20$, and 119 snapshots between $z=20$ and $z=0$, following the formula: $\log_{10}(1 + z_{N}) = N (N + 70) / 16800$ where $0 \leq N \leq 118$.

Friends-of-friends~\citep[\fof;][]{FoF} halos are identified at each snapshot using a linking-length of $0.2$ times the mean inter-particle separation. 
Subhalos and merger trees are constructed with \hbt~\citep{hbt,hbtplus}, a time-domain subhalo finder and merger tree builder that tracks the evolution of every halo in the simulation from the earliest to the latest snapshots. The \hbt algorithm identifies a particle group as a halo if it contains more than 20 particles. A halo is converted to a subhalo tracked by its self-bound particles once it is accreted onto another larger halo. A unique track ID is created to identify each halo throughout its evolutionary history including the subhalo phase, facilitating the straightforward retrieval of the evolution history. \hbt improves over conventional subhalo finders primarily in a more robust identification of subhalo in the inner halo where conventional finders suffer from obscuration effect, and in more consistent linking of subhalos across snapshots while conventional finders suffer from flip-flop of central-satellite membership and fragmentation of tree branch due to missing links~\citep[see][]{hbtplus,hbt,Srisawat2013MNRAS.436..150S,Onions2013MNRAS.429.2739O,2015MNRAS.454.3020B}. 
These improvements result in a more consistent and robust tree, which can affect the properties of individual galaxies evolved by the semi-analytical model~\citep[e.g.,][]{SussingSAM,HBTShark} and the abundances of galaxies due to the difference in the number of branches resolved~\citep{hbtplus}. 
Besides the \hbt catalogs, an alternative subhalo catalog built with the \subfind~\citep{subfind} code is also available for the primary runs, which are however not used in the current work. 

\begin{table*}[t]
\centering
\begin{threeparttable}
\footnotesize
\caption{Specifications of the simulations used in the current work. Box size and softening are in comoving units.}
\label{tab:primary}
\begin{tabular}{ccccccc}
\toprule
\hline
         &  Boxsize ($\mpch$) & Softening ($\kpch$) &  Particle Mass ($\msunh$) & Particles &  Snapshots & Cosmology \\\hline
         Jiutian-1G & 1000 & 4.0 & 3.723 $\times 10^8$  & $6144^3$  & 128 & Planck 2018  \\
 Jiutian-2G & 2000 & 7.0 & 2.978 $\times 10^9$ & $6144^3$  & 128 & Planck 2018 \\\hline
 \bottomrule
 \end{tabular}
 \end{threeparttable}
\end{table*}

\section{Mock Building Pipeline}
\label{sec:pipe}

\subsection{Semi-Analytical Galaxy Formation Model}
\label{sec:sam}

We use the semi-analytic code GAlaxy Evolution and Assembly (\gaea) \footnote{\url{https://sites.google.com/inaf.it/gaea/}} \citep{xie2020} to model galaxies from the available subhalo catalogs. The model builds on the semi-analytical model described in \citet{Gabriella2007MNRAS} and has been developed mainly on (i) chemical enrichment and non-instantaneous stellar recycling \citep{delucia2014}; (ii) redshift-dependent stellar feedback \citep{hirschmann2016}; (iii) H$_2$-based star formation law \citep{xie2017}; (iv) gradual strangulation and ram pressure stripping of cold gas for satellite galaxies \citep{xie2020}. In this work, we calibrate the model with a primary focus on the observed stellar mass function at $z=0$ from \citet{Li2009MNRAS}.  We tune the parameters in one sub-volume and then applied to the entire simulation box. We have also considered the HI mass function, HI mass-stellar mass relation, the black hole mass-bulge mass relation, quenched fractions at $z=0$ with lower weights. However, due to the high computational cost, these additional statistics are largely taken for reference without excessive optimization towards them. Table~\ref{tab:gaea_par} lists the key parameters for the two simulations. Parameters that are not listed here follow the values adopted in \citet{xie2020}.

\begin{table*}
\centering
\caption{Key \gaea parameters tuned for the Jiutian simulations. Impacts of these parameters and detailed descriptions are discussed in \citet[][and references therein]{DeLucia2024A&A}. }
\begin{tabular}{llcc}
\toprule
\hline
      Parameter & meaning &  Jiutian-1G & Jiutian-2G \\\hline
         $\epsilon_\mathrm{reheat}$ & Reheating efficiency &  0.5 &  0.7 \\
         $\epsilon_\mathrm{eject}$ & Ejection efficiency &  0.2    &  0.2 \\
         $\gamma_\mathrm{reinc}$ & Reincorporation efficiency &  1.0 & 1.0  \\
         $\kappa_\mathrm{radio}/10^{-5}$ & Hot gas black hole accretion efficiency &  1.0 &  5.0 \\
         $f_{\rm BH}$ & Cold gas accretion efficiency &  0.005 &  0.08 \\
         $\xi_{\rm slow}$ &  \makecell[l]{Ratio between specific angular momentum of
gas cooling \\ through "slow mode" and that of the halo} & 1.4 & 1.0 \\
         $\xi_{\rm rapid}$ & \makecell[l]{Ratio between specific angular momentum of
gas cooling \\ through "rapid mode" and that of the halo} & 2.0 & 1.4 \\
    \hline
 \bottomrule
\end{tabular}\label{tab:gaea_par}
\end{table*}

\gaea has been extensively modified to work with the data structure of the \hbt tree which uses persistent \texttt{TrackIds} instead of progenitor-descendant links to describe a tree. The code has also been modified to incorporate new features available in the \hbt tree. It has been re-calibrated to account for different merger rates recorded by different tree builders. 
In particular, the \hbt merger trees identify mergers between substructures, which allow \gaea to trace mergers between satellite galaxies. Compared with models that consider only central-satellite mergers, dealing with satellite-satellite mergers may decrease the merger rate of central galaxies, suppress the growth of SMBH, and consequently influence the strength of AGN feedback. 

The surviving galaxy during satellite-satellite merger is associated with the ID of the surviving subhalo, \texttt{SinkTrackId}, provided in the \hbt merger trees. If a subhalo is destroyed before merging into another subhalo, i.e., it has only 1 DM particle left in the \hbt catalog but without a valid \texttt{SinkTrackId}, the galaxy in it is labeled as an orphan galaxy in \gaea. We follow \citet{Gabriella2007MNRAS} to estimate the merger time-scale that is defined as the time since the loss of the subhalo until galaxy merger. 
Since the merging time scale estimated from dynamical friction depends on stellar mass, the estimated merging time-scales of orphan low-mass galaxies are much longer than those of more massive ones. In some cases, the time-scale for a low-mass satellite galaxy merging into a more massive satellite is much longer than the time scale for the massive satellite merging into the FoF central galaxy. 
We merge both satellite galaxies to the central galaxy in this situation.

The modified \gaea code processes the subhalo catalogs snapshot by snapshot, evolving all the galaxies in the snapshot at each time step. To speed up the computation, we split the merger trees of the entire simulation into mutually independent merger forests, each of which contains a collection of merger trees that are connected with some other trees in the same forest at some point in history through subhalo exchange. This enables parallel processing of different forests by \gaea. In subsequent parts of this paper, while referring to the properties of galaxies, we will use the term 'Jiutian galaxies' to indicate galaxies from the \gaea model run on the Jiutian $N$-body simulation.

\subsection{SED Generator: \starduster}
\label{sec:sed}

To allow for the outputs from our semi-analytical model (SAM) to be easily adapted for different surveys having different filters, we do not compute the broadband galaxy magnitudes on the fly, but only output their star formation histories (SFHs) at each snapshot. Each SFH\footnote{The star formation history used here contains four parts: stellar disk formation history, stellar bulge formation history, metal disk formation history, metal bulge formation history.} is then processed with \starduster\footnote{\url{https://github.com/yqiuu/starduster}}\citep{2022ApJ...930...66Q} to generate full Spectral Energy Distributions (SEDs), which is further filtered to produce magnitudes in various bands.

\starduster is a Python package for synthesizing galaxy SEDs based on an input SFH and a stellar emission library. By default, it employs the Flexible Stellar Population Synthesis~\citep[FSPS;][]{Conroy2009ApJ,Conroy2010ApJ} model with a Chabrier initial mass function \cite{Chabrier2003PASP} to construct intrinsic stellar spectra. The distinct advantage of \starduster lies in its efficient and flexible treatment of dust attenuation and emission. It employs a neural network trained on radiative transfer simulations run with the \textsc{SKIRT} code\cite{Baes2011ApJS..196...22B,Camps2015A&C.....9...20C,Camps2020A&C....3100381C}, which allows for adjustable geometry configurations.
For comprehensive implementation details of \starduster, we direct readers to the original methodology paper \citep{2022ApJ...930...66Q}. Below we briefly describe the key galaxy geometry assumptions and our modifications to the dust attenuation treatment when combining \starduster with \gaea outputs.


\subsubsection{Galaxy Geometry Model}

Dust attenuation can significantly alter the SED of actively star-forming galaxies which are known to be dust-rich. To model this, \starduster describes each galaxy with three relevant components for the SED calculation, including an exponential stellar disk, a Sérsic-profile bulge, and an associated dust disk. The corresponding density profiles can be described in cylindrical coordinates as
\begin{equation}
    \rho_\mathrm{disk}(r, z) \propto \mathrm{exp}(-\frac{r}{r_\mathrm{disk}} - \frac{|z|}{h_\mathrm{disk}}), 
\end{equation}

\begin{equation}
    \rho_\mathrm{bulge}(r, z) \propto {S}_n(\frac{\sqrt{r^2+z^2}}{r_\mathrm{bulge}}),
\end{equation}
and 
\begin{equation}
    \rho_\mathrm{dust}(r, z) \propto \mathrm{exp}(-\frac{r}{r_\mathrm{dust}} - \frac{|z|}{h_\mathrm{dust}}),
\end{equation}
where $r_\mathrm{disk}$, $r_{\rm bulge}$, $r_{\rm dust}$ specify the scale radius of each component, and $h_\mathrm{disk}$ and $h_{\rm dust}$ are the scale heights.
The Sérsic function ${S}_n$  is defined as the Abel inversion,
\begin{equation}
    S_n(s) = -\frac{1}{\pi} \int_s^{\infty}\frac{dI}{dt} \frac{dt}{\sqrt{t^2-s^2}},
\end{equation}
of the projected Sérsic profile,
\begin{equation}
    I(t) = \frac{b^{2n}}{\pi\Gamma(2n+1)}\mathrm{exp}(-bt^{1/n}),
\end{equation} where $b$ is a dimensionless parameter that depends on the Sérsic index $n$, and $\Gamma(n)$ is the Gamma function~\cite{Baes2011A&A}.

The \gaea model directly predicts the stellar disk and bulge parameters of each galaxy, in addition to parameters describing a gas disk.  
In this work, we set the dust disk radius to 0.6 times that of the gas disk, and find that the resulting model can well predict the dust-attenuated luminosity function from $u$ to $z$ bands at redshift 0. In \starduster, the height of the dust disk is fixed to be $0.1$ times that of the stellar disk. The total mass of the dust disk is derived from the metal gas mass of the galaxy, $M_{\rm dust}=0.33 M_{\rm Z}$ \citep{Ruyer2014A&A, Lagos2019MNRAS}.

The attenuation curve also depends on the inclination of the disk, in addition to the geometry parameters above. For each galaxy, we assign a random inclination, $\theta$, uniformly distributed in $\cos\theta$ between 0 and 1.

\subsubsection{Interstellar Medium (ISM) Extinction}

For each galaxy with stellar and dust components specified above, \starduster can predict the attenuation by the dust disk on the stellar disk and bulge separately, according to the neural network model trained by using the \textsc{skirt} simulation data.
The composite attenuated luminosity is given by 
\begin{align}
L_{\lambda, \mathrm{galaxy}} &= 10^{-0.4A_{\lambda, \mathrm{disk}}}L_{\lambda, \mathrm{disk}} + 10^{-0.4A_{\lambda, \mathrm{bulge}}}L_{\lambda, \mathrm{bulge}},
\end{align}
where $L_{\lambda, \mathrm{disk}}$ and $L_{\lambda, \mathrm{bulge}}$ denote intrinsic stellar luminosities, while $A_{\lambda, \mathrm{disk}}$ and $A_{\lambda, \mathrm{bulge}}$ represent component-specific attenuation curves predicted by \starduster.  The attenuation can be alternatively expressed in terms of the optical depth, $\tau_{\lambda}$, of the ISM dust through the transmission function,
\begin{equation}
T \equiv \frac{I_\mathrm{out}}{I_\mathrm{in}} = e^{-\tau_{\lambda}} = 10^{-0.4A_{\lambda}}.
\label{equ:T}
\end{equation}

\subsubsection{Molecular Birth Cloud Extinction}

Young stellar populations (with ages $<10 \mathrm{Myr}$) experience additional attenuation from molecular birth clouds\cite{Charlot2000ApJ}, a feature absent in the original \starduster implementation. Following papers like \cite{Kong2004MNRAS,Gabriella2007MNRAS,Henriques2015MNRAS}, we relate the birth cloud optical depth $\tau_\mathrm{BC}$ to the ISM optical depth $\tau_\mathrm{ISM}$ through,
\begin{equation}
    \tau_{\lambda, \mathrm{BC}} = \tau_{\lambda, \mathrm{ISM}}(\frac{1}{\mu}-1)(\frac{\lambda}{5500\textnormal{\AA}})^{-0.7},
\label{equ:ism2bc}
\end{equation}
where $\mu$ follows a truncated Gaussian distribution with mean 0.5 and standard deviation 0.2, truncated at 0.1 and 1. $\tau_{\lambda, \mathrm{ISM}}$ for each galaxy is the output from the deep learning model of \starduster, for the disk and bulge components separately.

Combining the attenuation of the ISM and the birth cloud, the total optical depth experienced by a stellar population of age, $t$, is given by
\begin{equation}
  \tau_{\rm tot} =\begin{cases}
    \tau_{\rm ISM}+\tau_{\rm BC}, & \text{if $t<10$ Myr}.\\
     \tau_{\rm ISM}, & \text{otherwise}.
  \end{cases}
\end{equation}

\begin{figure*}[ht]
\centering
    \includegraphics[width=0.45\linewidth]{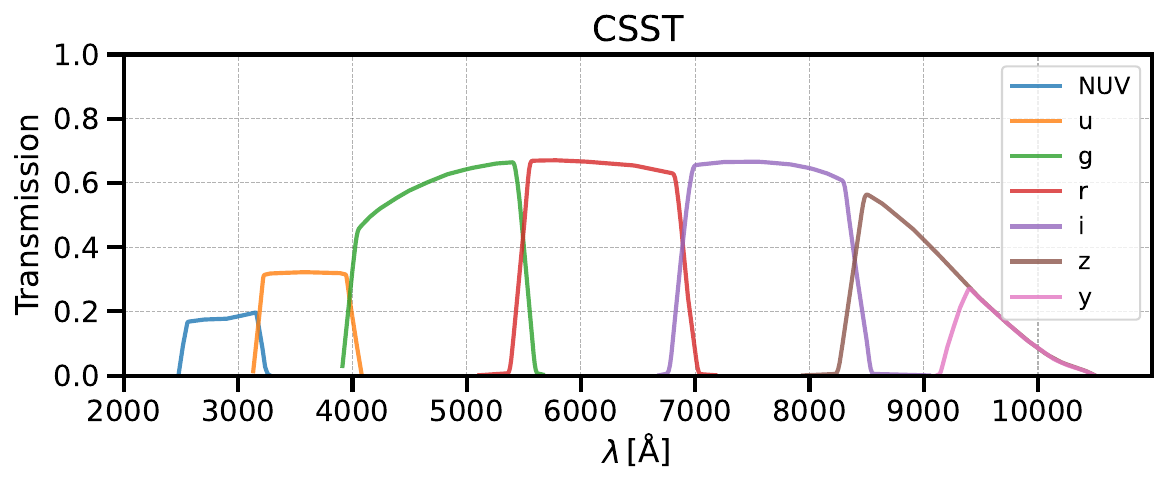}
    \includegraphics[width=0.45\linewidth]{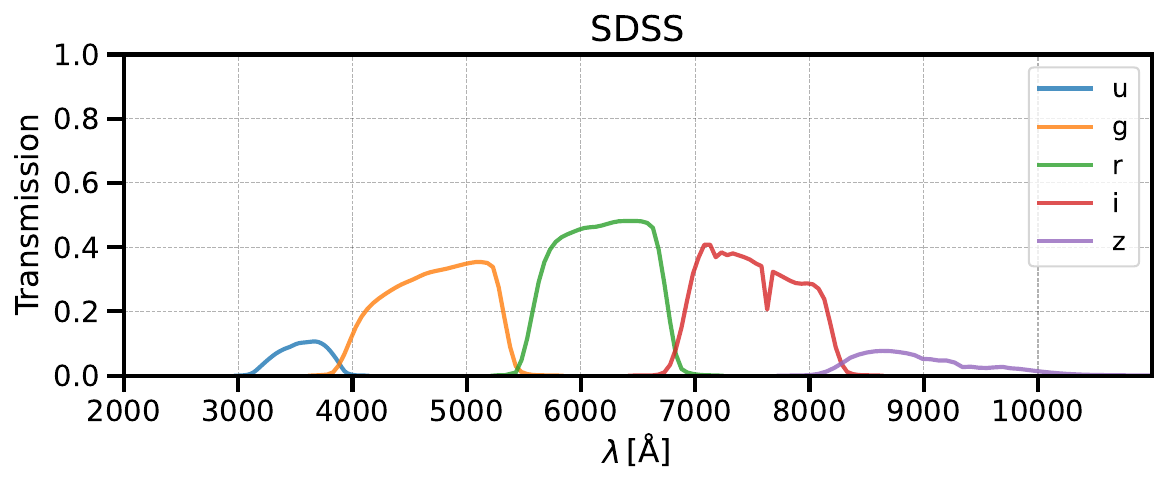}
    \includegraphics[width=0.45\linewidth]{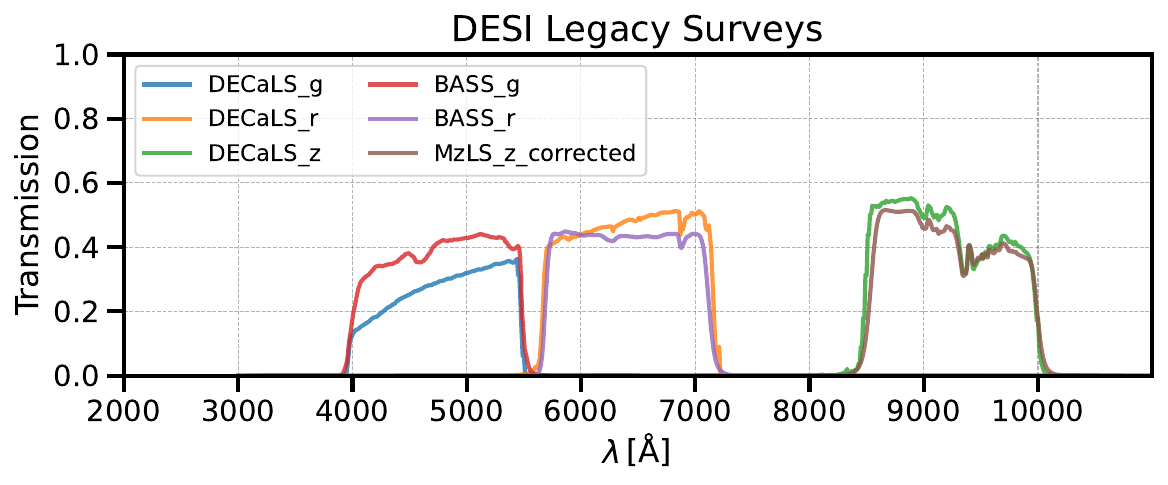}
    \includegraphics[width=0.45\linewidth]{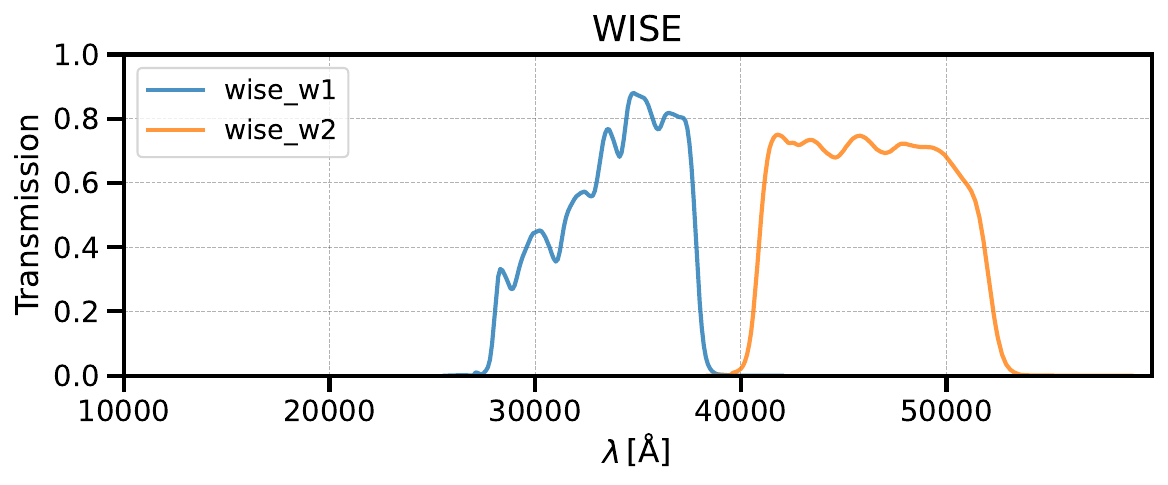}
    \caption{Transmission curves of various photometric filters adopted in this work, as labeled.}
    \label{fig:filters}
\end{figure*}

\subsubsection{Filter Systems}
\label{subsec:filters}

While the full spectral energy distributions (SEDs) are computed for all galaxies, it is too expensive to store the high resolution SED for each galaxy at each redshift. Instead, we compile a list of 20 filters from four major surveys, and only store the magnitudes through these filters (see Figure~\ref{fig:filters}), including both dust-processed and intrinsic magnitudes. 
For CSST, our adopted transmission curves incorporate the full system response including the filter transmission, detector quantum efficiency and the mirror reflection\footnote{\url{https://svo2.cab.inta-csic.es/theory/fps/index.php?mode=browse&gname=CSST&gname2=SC0-Phot&asttype=}}.
We also incorporate filters from three constituent surveys of the Dark Energy Spectroscopic Instrument Legacy Imaging Surveys (DESI-LS)~\cite{DESI_LS}, namely the \textit{g}, \textit{r}, \textit{z} bands from the Dark Energy Camera Legacy Survey (DECaLS), the \textit{g} and \textit{r} bands from the Beijing–Arizona Sky Survey (BASS) and the Mayall \textit{z}-band Legacy Survey (MzLS). The filter response definition takes into account instrumental corrections for telescope throughput, optical corrector transmission, and atmospheric effects\footnote{\url{https://www.legacysurvey.org/dr8/description/}}\cite{DESI_LS}.
Additionally, we include two mid-infrared bands from the Wide-field Infrared Survey Explorer (WISE) \citep{2010AJ....140.1868W}: \textit{W1} (3.4 µm) and \textit{W2} (4.6 µm), critical for tracing dust emission processes.

In the AB magnitude system, the magnitudes from different filters are insensitive to the detailed filter response curves. For example, the magnitudes through the CSST filters typically differ from those under the same SDSS filters by less than 0.2 magnitudes for $z<1.7$. So we only include magnitudes from the CSST and WISE filters to reduce the amount of public data release.

\subsection{Light-Cone Generator: \blic}
\label{sec:lc}

In this section, we introduce our Python-based light-cone construction tool \blic \footnote{\url{https://github.com/zltan000/BLiC}} for generating mock observational catalogues from numerical simulations. The package converts discrete simulation snapshots into continuous light-cone observations while accounting for survey geometry and observational effects. It implements features like random tiling strategies~\citep{2005MNRAS.360..159B} and multiple temporal interpolation schemes to reconstruct galaxy motions between snapshots. 

The light-cone construction process begins by setting up the observer's coordinates and survey parameters. Our algorithm adaptively tiles simulation boxes along the line of sight, and only processes boxes intersecting the observer's light-cone, significantly lowering the computational cost. 
For each object in a relevant box, we solve for its observed redshift in the light-cone according to the intersection condition,
\begin{equation}
    D_{\mathrm{comoving}}(z)\equiv \int_{0}^{z}\frac{c\mathrm{d}z'}{H_0 \sqrt{\Omega_{m}(1+z')^3+\Omega_{\Lambda}}}= \left \| \vec{X}(z) \right \|,
    \label{eq1}
\end{equation}
where $c$ is the speed of light, and $H_0$, $\Omega_m$, and $\Omega_\Lambda$ are the Hubble constant, matter, and dark energy density parameters at $z=0$. 

\begin{figure}[H]
\centering
    \includegraphics[width=0.48\textwidth]{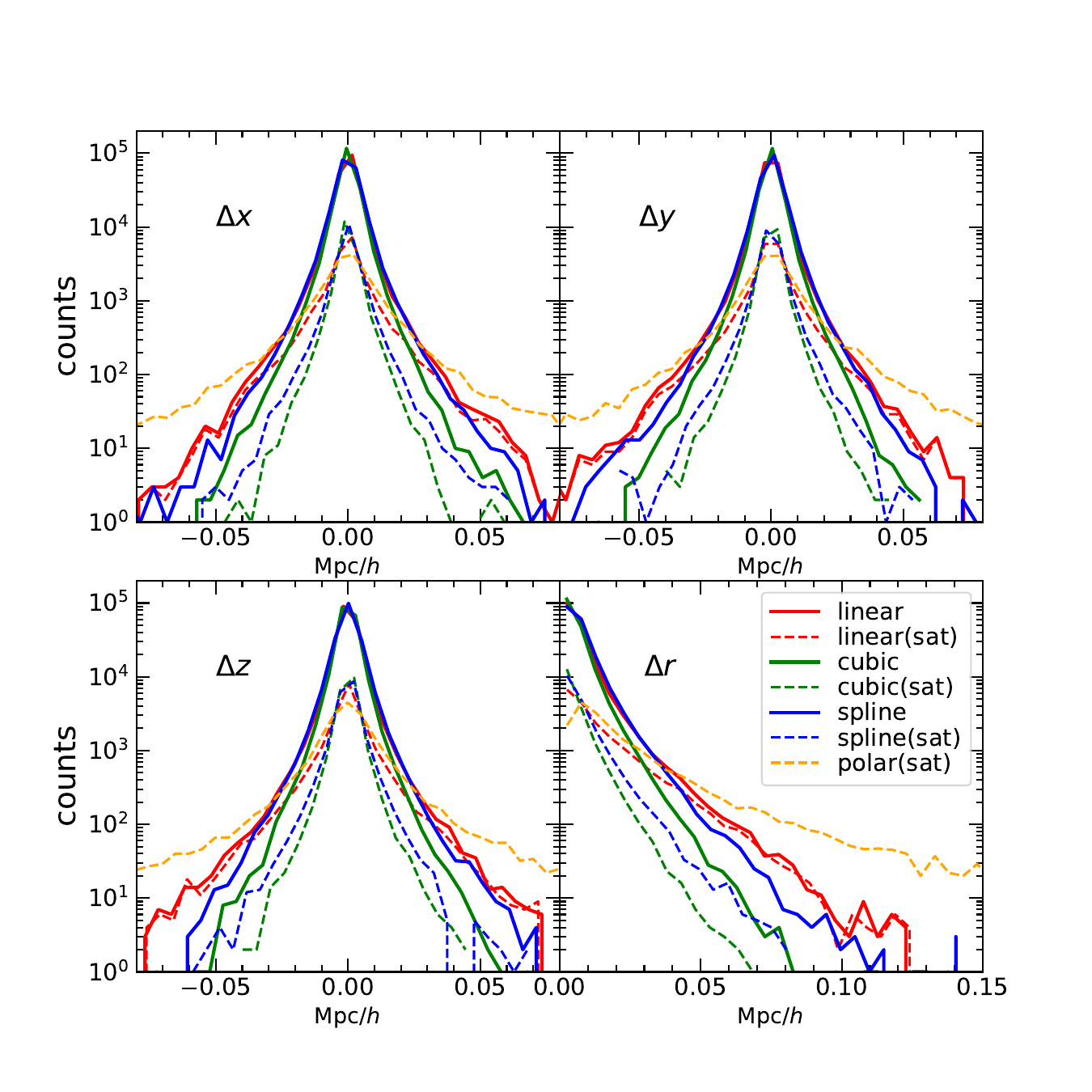}
    \caption{Accuracy in interpolating the galaxy position. Different interpolation schemes (shown with different colors as labeled) are adopted to interpolate the trajectory of each galaxy in the Illustris-3 simulation, using snapshots both before and after the 68th snapshot, with a time separation of $0.3$ Gyr between them. The interpolation results at the time of the 68th snapshot is compared against the snapshot itself, and the distribution of the residuals in the predicted coordinates are shown for each of the $(x,y,z)$ dimension. The bottom right panel shows the total displacement, $\Delta r=\sqrt{\Delta x^2+\Delta y^2 +\Delta z^2}$.
    Solid lines show the distribution of all galaxies, while the dashed lines show the results for satellite galaxies only.
   }
    \label{fig:pos_predict_comp}
\end{figure}

An accurate solution to the above equation requires a continuous trajectory of the object, $\vec{X}(z)$, which can only be obtained through interpolation of the trajectory recorded at discrete snapshots. We implement a few different interpolation schemes in \blic, including linear, cubic, and cubic spline interpolations, as well as linear interpolation in polar coordinates. The interpolations mainly use the positions of the object in adjacent snapshots to solve for interpolation coefficients, except for cubic interpolation, which also makes use of the velocities. In Supplementary materials, we provide detailed descriptions of the interpolation schemes as well as some tests on their performance. Figure~\ref{fig:pos_predict_comp} shows one example test comparing the accuracy of different interpolations in predicting the positions of galaxies in the Illustris-3 simulation\footnote{Here, we employ an HBT-converted version of the Illustris-3 simulation. Its configuration—including the definition of positions, the method for linking merger trees, and the data formats—is consistent with the Jiutian simulation. We selected this smaller simulation box to validate our schemes with a reduced computational burden, given the large box size of the Jiutian simulation suites.}~\cite{Vogelsberger2014Natur, Genel2014MNRAS, Vogelsberger2014MNRAS}. For this test, we have used galaxies from snapshots both before and after the 68th snapshot in Illustris-3, and interpolate the trajectories to snapshot 68 to compare against the result in the snapshot itself. 
As can be seen in Figure~\ref{fig:pos_predict_comp}, cubic interpolation produces the highest accuracy in predicting the galaxy positions. We thus adopt cubic interpolation as our default scheme. 

Figure~\ref{fig:pos_predict_comp} also shows that the linear interpolation in polar coordinates provides the least accurate reconstruction of the displacements. This appears to contradict \citet{2013MNRAS.429..556M}, who adopted polar interpolation as their optimal scheme. We also note that their choice focuses on optimizing the correlation function prediction. Indeed, in  Supplementary materials we show that polar interpolation does not perform significantly worse than other schemes in predicting the correlation function.

Once the coordinates and redshift for each object in the light-cone have been fixed, we further interpolate additional physical quantities of model galaxies at the same cosmic epoch, such as stellar mass and luminosity. This ensures that the resulting light-cone catalog provides the correct information for a continuous distribution of galaxies. In the current version of \blic, only linear and nearest neighbour interpolations are implemented for interpolating the physical properties.

\section{Galaxy Population Properties}
\label{sec:results}

This section compares some key statistics from our mock galaxy catalogs against observations, including the stellar mass function (SMF), luminosity function, gas fractions, size-mass relations from simulation boxes, and galaxy clustering from the light-cone.

\subsection{Stellar Mass Function and Resolution Limit}

Stellar mass serves as a fundamental property to trace galaxy evolution and is strongly related with derived parameters like magnitudes in theoretical models. As a primary output of semi-analytical and hydrodynamical models, it remains a key proxy for model calibration \citep[e.g.,][]{Katsianis2021ApJ, Xie2023SCPMA..6629513X}.

\begin{figure}[H]
    \includegraphics[width=0.48\textwidth]{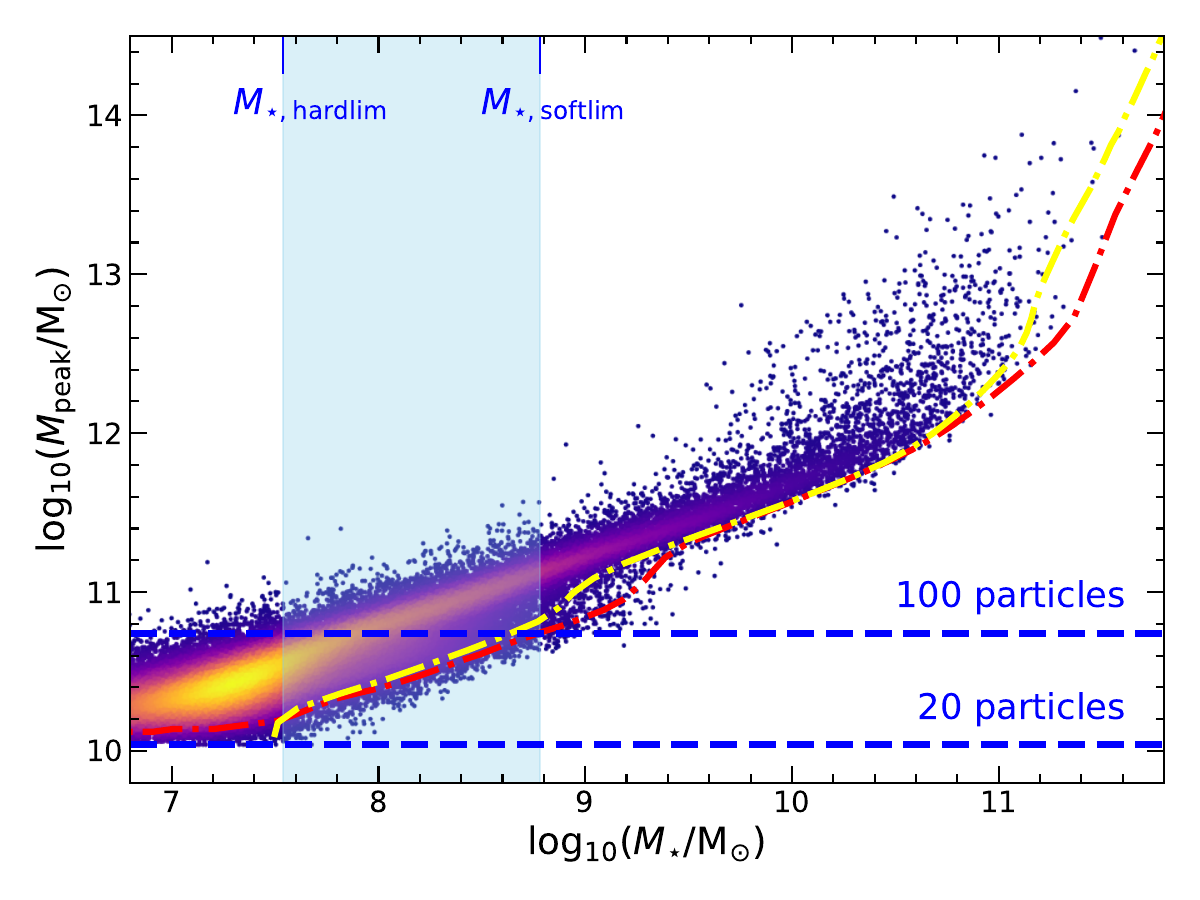}
    \caption{Subhalo peak mass versus stellar mass for Jiutian-1G galaxies. The red dash-dotted curve shows the 96\% lower bound of the peak mass at a fixed stellar mass. At a stellar mass of $M_{\rm \star, softlim}=10^{8.74} \msun$, this peak limit reaches 100 particles as marked by the blue dashed horizontal line. The yellow dash-dotted curve shows the 96\% upper limit of the stellar mass at a fixed peak mass. For a peak mass of 20 particles, this stellar mass limit reaches $M_{\rm \star, hardlim}=10^{7.50} \msun$. The vertical blue band marks the mass range between the soft and hard stellar mass limits. }
    \label{fig:mp_ms1G}
\end{figure}

In computing stellar mass statistics from simulations, it is important to realize that the galaxy population becomes incomplete below a certain mass limit due to the finite resolution of the simulation. To find out this limit, in Figure~\ref{fig:mp_ms1G} we show the stellar mass-subhalo peak mass relation of our model galaxies in Jiutian-1G. As the stellar mass largely stops growth once a galaxy becomes a satellite~\cite{Wang2006MNRAS,Wang2010MNRAS,Watson2013ApJ}, it is highly correlated with the peak mass of a subhalo, which represents the maximum subhalo mass before it becomes a satellite. Thus in order to ensure completeness in stellar mass statistics, we can limit our galaxy sample to those residing in well resolved subhalos above a certain peak mass. Although the peak mass function from \hbt is expected to be complete down to 20 particle mass~\cite{hbtplus}, we pick 100 dark matter particle mass as a conservative mass limit for the peak mass. Above this limit, the structure of a subhalo can be reasonably well resolved so that the predicted galaxy properties, including the stellar mass and other properties such as disk size, are more reliable. We will call this a soft limit. As shown in Figure~\ref{fig:mp_ms1G}, at a stellar mass of $M_{\rm \star, softlim}=10^{8.74}\ \mathrm{M_\odot}$, almost all (96\%) of the galaxies have a peak mass above 100 particle mass. Thus we can use this stellar mass limit as a completeness limit, below which resolution effects may start to contaminate the stellar mass statistics. For Jiutian-2G, the corresponding soft limit is $M_{\rm\star,softlim}=10^{10.12}\ \mathrm{M_\odot}$.

An alternative but more aggressive limit can be obtained by examining the stellar mass of the lowest mass subhalo with 20 dark matter particles. According to the yellow curve in Figure~\ref{fig:mp_ms1G}, at the lowest peak mass, almost all (96\%) of the galaxies have a stellar mass below $10^{7.50} \msun$ in Jiutian-1G. Below this limit, the galaxy population can have large contributions from under-resolved or unresolved subhalos with 20 or less dark matter particles. We call this a hard limit. For Jiutian-2G, the corresponding limit is $M_{\rm\star,hardlim}=10^{9.15}\msun$. 

\begin{figure*}
    \centering
    \includegraphics[width=0.99\linewidth]{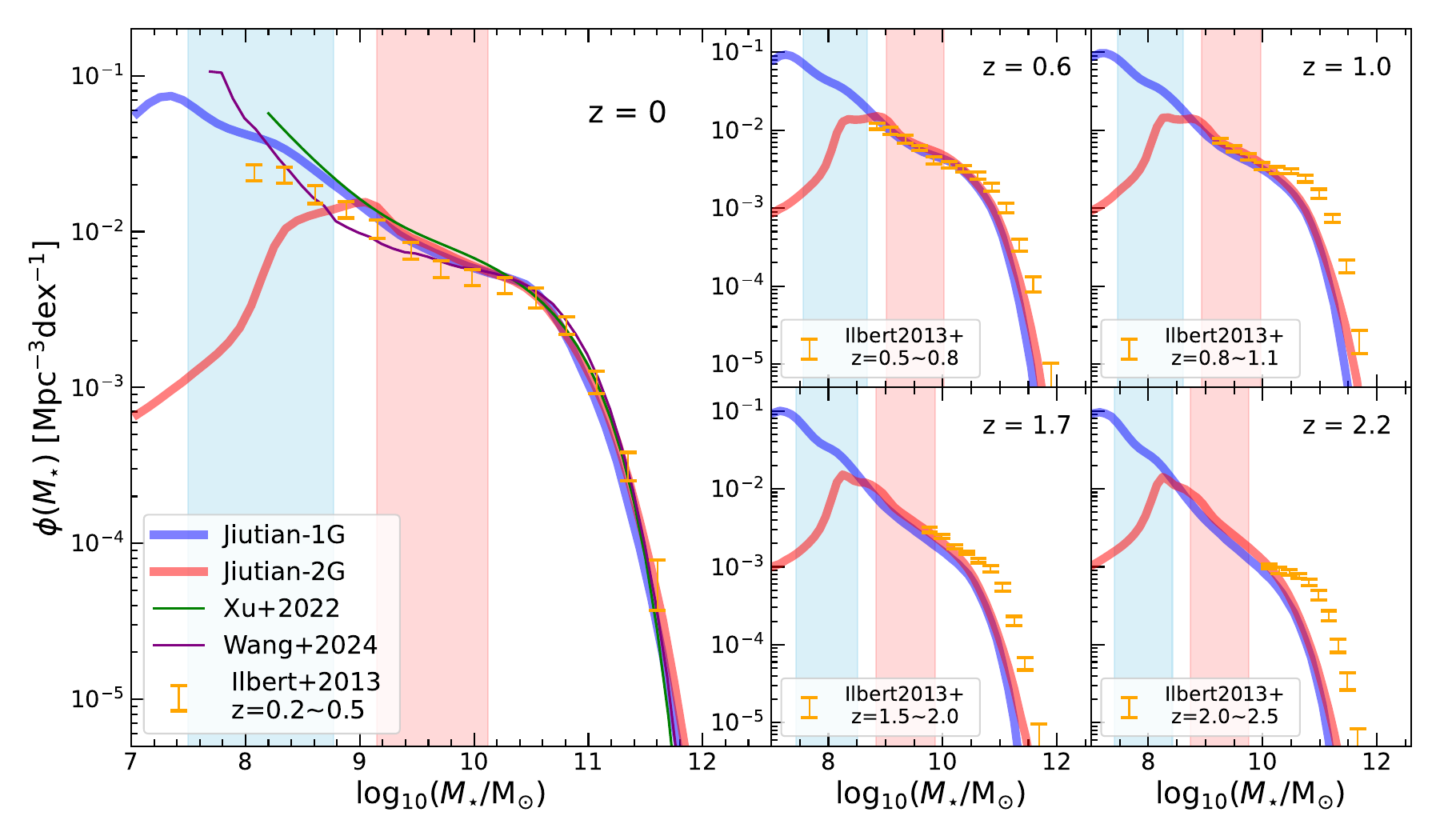}
    \caption{Stellar mass functions for Jiutian-1G (blue) and 2G (red) at multiple redshifts, compared with observational data~\cite{Ilbert2013A&A, Xu2022, WangYR2024}. For each simulation mass function, a vertical band of the corresponding color marks the range spanned between the hard and soft completeness limits as determined following Figure~\ref{fig:mp_ms1G} at the corresponding redshift. }
    \label{fig:sm}
\end{figure*}

These completeness limits can also be verified through convergence of the stellar mass functions (SMFs) with resolutions. In Figure~\ref{fig:sm} we show the SMFs for both Jiutian-1G and Jiutian-2G galaxies. As both models are calibrated against the observed SMF at $z=0$, they agree well with each other at $z=0$ by construction. At the low mass end, however, the two start to diverge, which is an indication of resolution effects. It can be seen that the convergence is well maintained above the soft limit, and that the Jiutian-2G SMF only deviates from the corresponding Jiutian-1G SMF at a mass below the hard limit of $M_{\rm\star,lim}=10^{9.15}\ \mathrm{M_\odot}$ derived above. 
These conclusions hold also when higher redshift SMFs are considered (see right hand panels in Figure~\ref{fig:sm}). We want to stress here that we have not calibrated the model to $z>0$ SMFs.

\begin{figure*}
    \centering
    \includegraphics[width=0.99\textwidth]{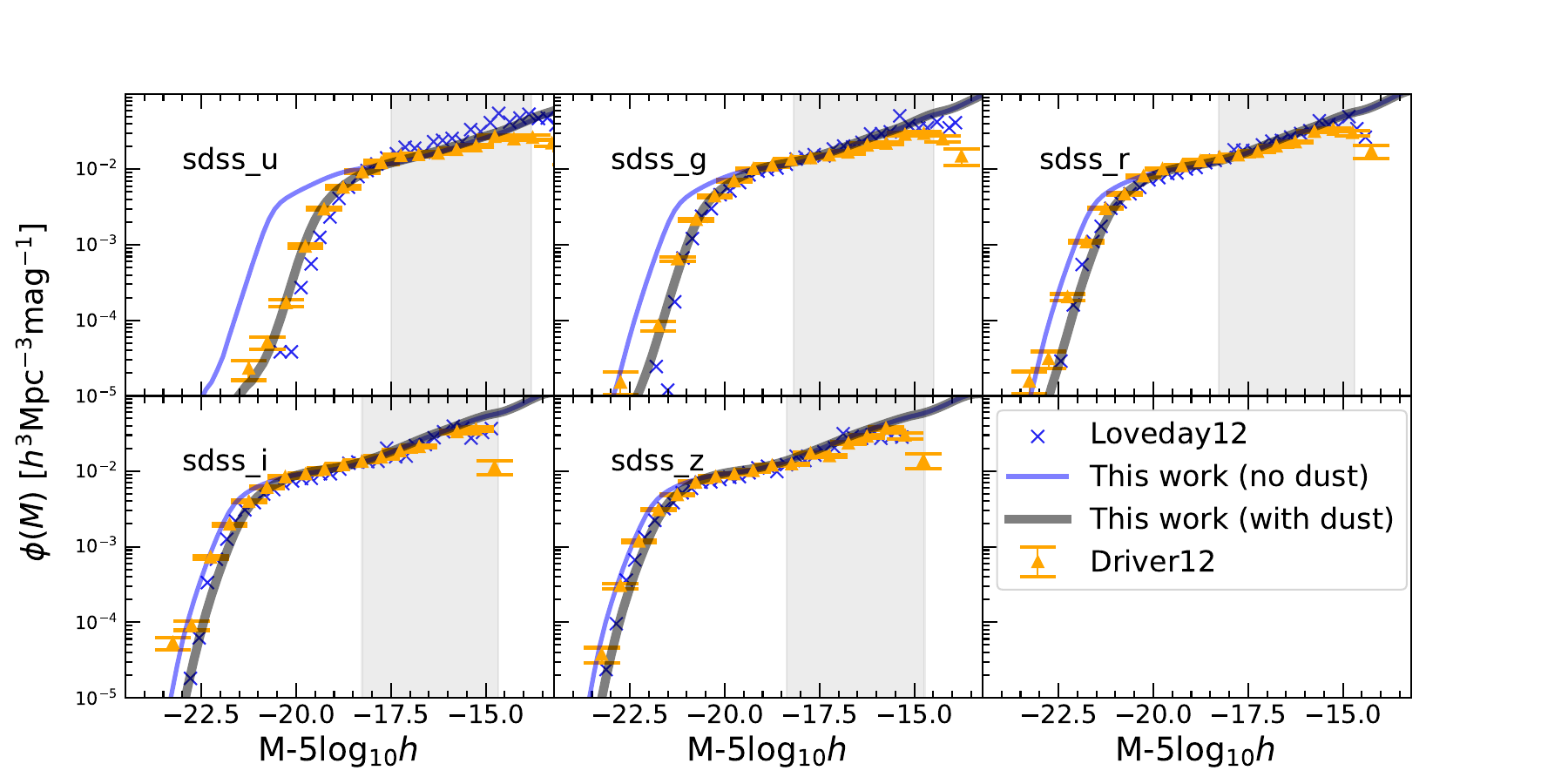}
    \caption{$z=0$ luminosity functions in SDSS bands of \gaea galaxies in Jiutian-1G. Blue curves show the distribution of intrinsic magnitudes, while gray curves include dust attenuation. In each panel, the vertical band span the hard and soft completeness limits in the modelled magnitudes.
    Observational measurements from the GAMA survey by \citet{Driver2012MNRAS} and \citet{Loveday2012MNRAS} are also shown by symbols as labelled.} 
    \label{fig:LF}
\end{figure*}

Figure~\ref{fig:sm} also shows some observational measurements of the SMFs at various redshifts. At $z=0$, the observational results at the low mass end show divergence among different works. Towards higher redshifts, the model SMFs tend to underpredict the observed SMFs at the high mass end, and the discrepancy increases with redshift. 
While current observational results at high redshifts carry inherent uncertainties related to sample selection and mass estimation techniques, this deficit stems primarily from the delayed star formation history for massive galaxies in our adopted model of \citet{xie2020}. The model under-predicts the passive populations during critical cosmic epochs due to the known limitations in the AGN feedback at high redshifts~\citep{DeLucia2024A&A}, leading to excessive star formation at later times.  
In a recent series of papers \cite{DeLucia2024A&A,Xie2024ApJ,Fontanot2025A&A}, we have presented a new version of the \gaea code, which provides a better agreement with the statistics of quenched galaxies. With respect to the model used in this paper, the new model also includes a more realistic treatment of AGN feedback. We plan to include this new \gaea version in future runs on the Jiutian simulation suite.

\subsection{Luminosity Functions}

Figure~\ref{fig:LF} presents the galaxy luminosity functions in five SDSS bands ($u$, $g$, $r$, $i$, $z$) from Jiutian-1G, compared with observations from the GAMA survey~\cite{Driver2012MNRAS,Loveday2012MNRAS}.
Dust attenuation causes substantial suppression of galaxy luminosities in these bands, especially in the $u$ band where the magnitudes can be reduced by $\sim1$~$\rm mag$. The completeness limits are also shown as the gray band in each panel, which spans the soft and hard limits, determined in a similar way as the stellar mass limits but in the peak mass-magnitude dimension. With dust extinction, the predicted luminosity functions agree well with the observations.
Note the observational measurements show some systematic differences due to the different magnitude systems adopted, with the results from \citet{Driver2012MNRAS} using Kron magnitudes systematically brighter than those from \citet{Loveday2012MNRAS} using Petrosian magnitudes.

\subsection{Gas Fractions and Size-Mass Relations}
\label{sec:gas_size}

\begin{figure*}[p]
    \centering
    \includegraphics[width=0.99\linewidth]{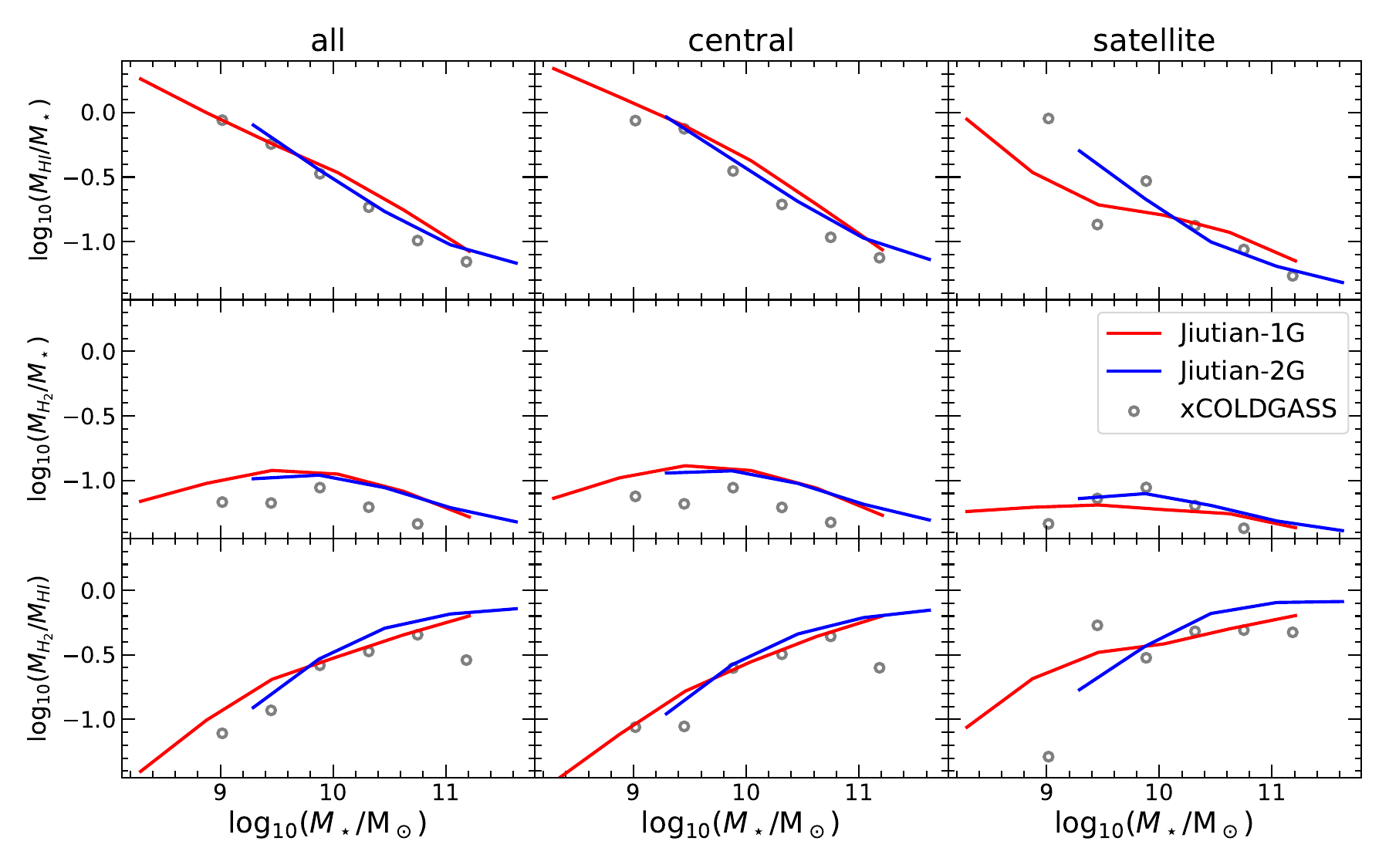}
    \caption{Gas phase properties versus stellar mass. Columns distinguish galaxy types. Rows show HI fractions (top), $\rm {H_2}$ fractions (middle), and molecular-to-atomic gas ratios (bottom). Gray circles represent xCOLDGASS observations combining HI measurements from \citet{Catinella2018MNRAS} and CO-derived $\rm {H_2}$ measurements from \citet{Saintonge2017ApJS}. The red and blue curves show predictions from the \gaea mocks of the two simulations as labelled.}
    \label{fig:HI_comp}
\end{figure*}

\begin{figure*}[p]
    \centering
    \includegraphics[width=0.99\linewidth]{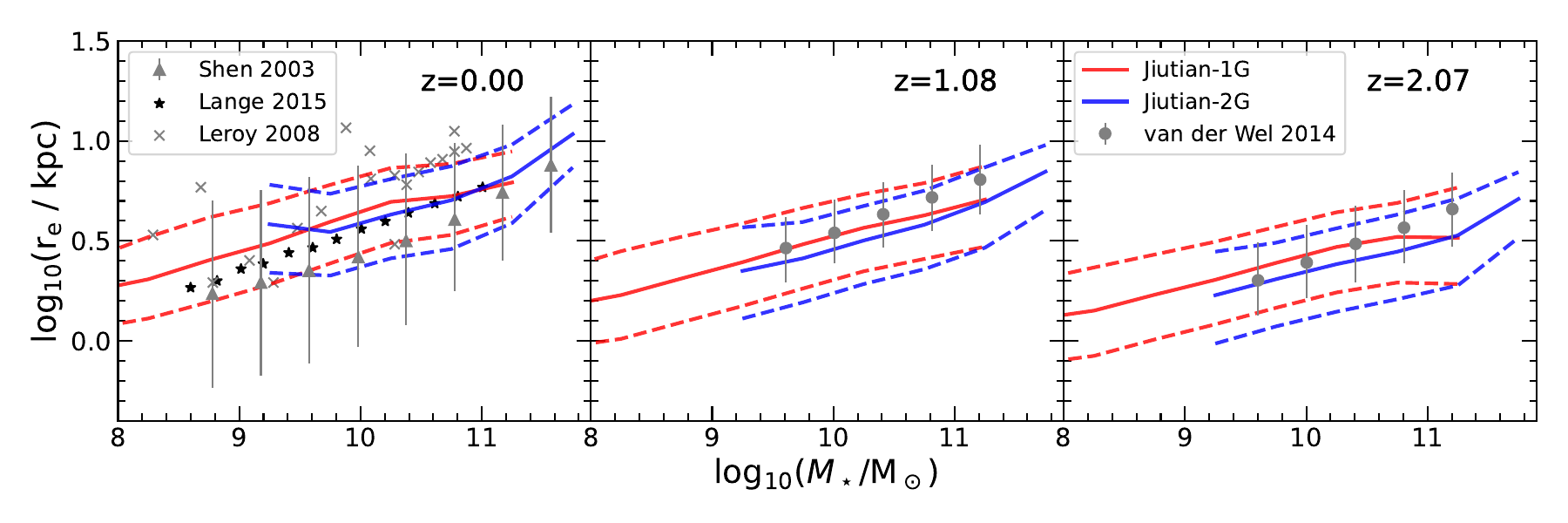}
    \caption{The stellar effective radius distributions in different stellar mass bins at different redshifts. Observational measurements from \citet{ShenSY2003MNRAS,Leroy2008AJ,Lange2015MNRAS} at $z=0$ and \citet{vanderWel2014ApJ} at $z>1$ are shown with symbols as labelled. 
    The solid lines are the median relations in the \gaea outputs, while the dashed lines enclose the 68\% scatter regions.}
    \label{fig:stellarsize}
\end{figure*}

One of the major strengths of the \gaea model lies in its success in predicting the galaxy size distribution and their atomic and molecular gas content.
Figure~\ref{fig:HI_comp} compares the molecular and atomic gas fractions in our mock catalogs with observational constraints from xGASS~\cite{Catinella2018MNRAS} and xCOLDGASS~\cite{Saintonge2017ApJS} (we simply refer to this combined sample as xCOLDGASS in the following text).
For fair comparisons, we apply the same $\mathrm{H_2}$ mass fraction cuts as in the observations (see Fig. 10 of \citet{Saintonge2017ApJS}), with $\mathrm{log}f_{{M_{\mathrm{H}_2}}/M_\star}>-1.6$ for galaxies with $10^9<M_\star<10^{10}\mathrm{M_\odot}$, and $\mathrm{log}f_{{M_{\mathrm{H}_2}}/M_\star}>-1.8$ for galaxies with $M_\star>10^{10}\mathrm{M_\odot}$.

Model predictions from the two simulations show good consistency across all mass ranges and galaxy types, indicating our modeling of gas physics is resolution independent. Both simulations also align well with observational data, especially in capturing the characteristic $\rm H_2$ fraction turnover at $M_\star \approx 10^{10}\rm M_\odot$. Satellite galaxies exhibit systematically lower gas fractions than centrals at fixed galaxy stellar mass, as a consequence of the tidal and ram-pressure stripping as they orbit within their host halos. 

Figure~\ref{fig:stellarsize} examines the stellar effective radii across redshifts. Again, predictions from Jiutian-1G and 2G show very good convergence across redshifts. At $z=0$, the measurements of \citet{Lange2015MNRAS} are very close to the prediction from our simulations. The results of \citet{ShenSY2003MNRAS} are also consistent with our predictions within the measurement uncertainty. We note that these two works measure the average size-mass relations. The measurements of \citet{Leroy2008AJ} for individual galaxies are also consistent with the predicted distribution of our models within the scatter that is indicated by the dashed curves. 
The mass-dependent slope evolves from $z=0$ to $z=2$, indicating accelerated size growth in massive galaxies at early cosmic epochs. At higher redshifts ($z=1$ middle panel; $z=2$ right panel), the simulations maintain good agreement with observational measurements from \citet{vanderWel2014ApJ}. These agreements highlight the success of \gaea in modeling the structural evolution of galaxies.

\subsection{Galaxy Clustering}

\begin{figure*}[htbp]
    \centering
    \includegraphics[width=0.99\linewidth]{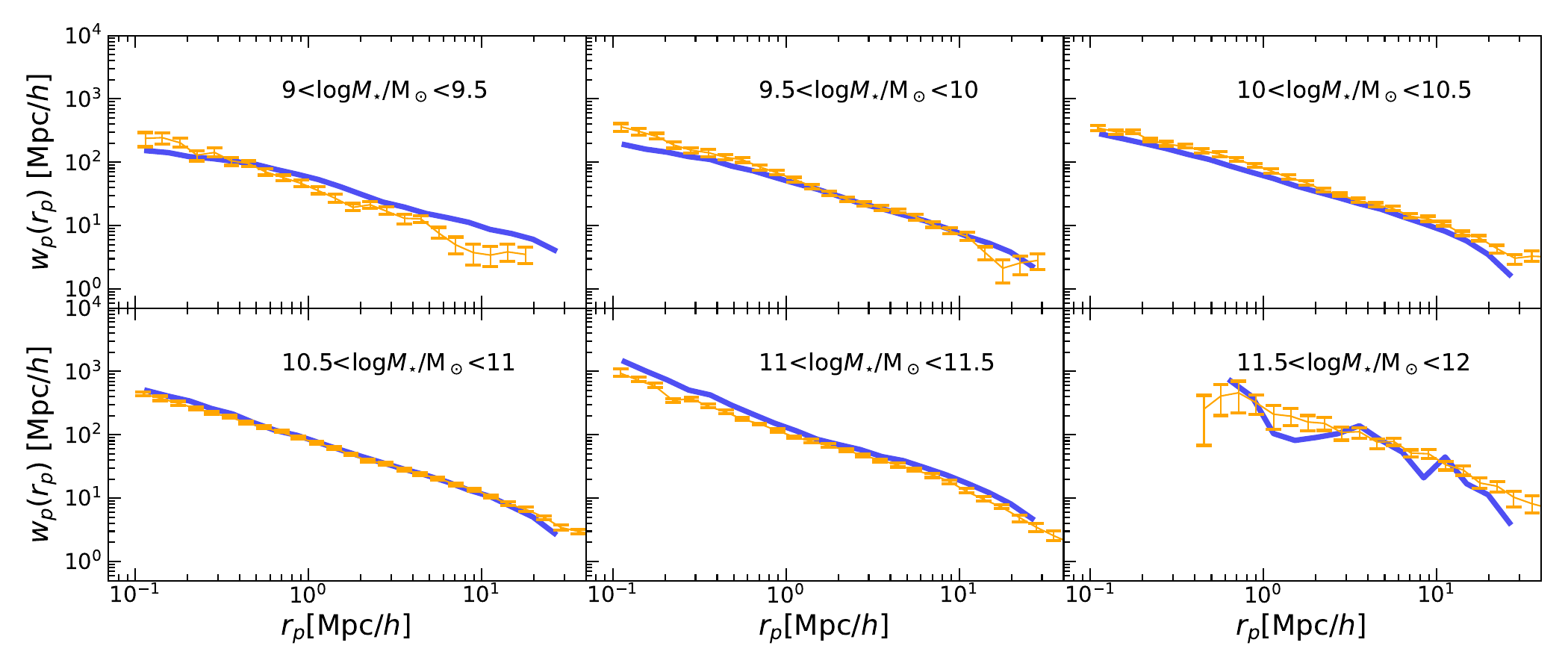}
    \caption{The projected correlation function, $w_p(r_p)$ across different stellar mass bins. Blue curves show results of our work, while orange curves with error bars represent observational measurements from the SDSS survey reported in \citet{Li2006}.}
    \label{fig:proj_corr}
\end{figure*}

\begin{figure*}[htbp]
\centering
    \includegraphics[width=0.48\linewidth]{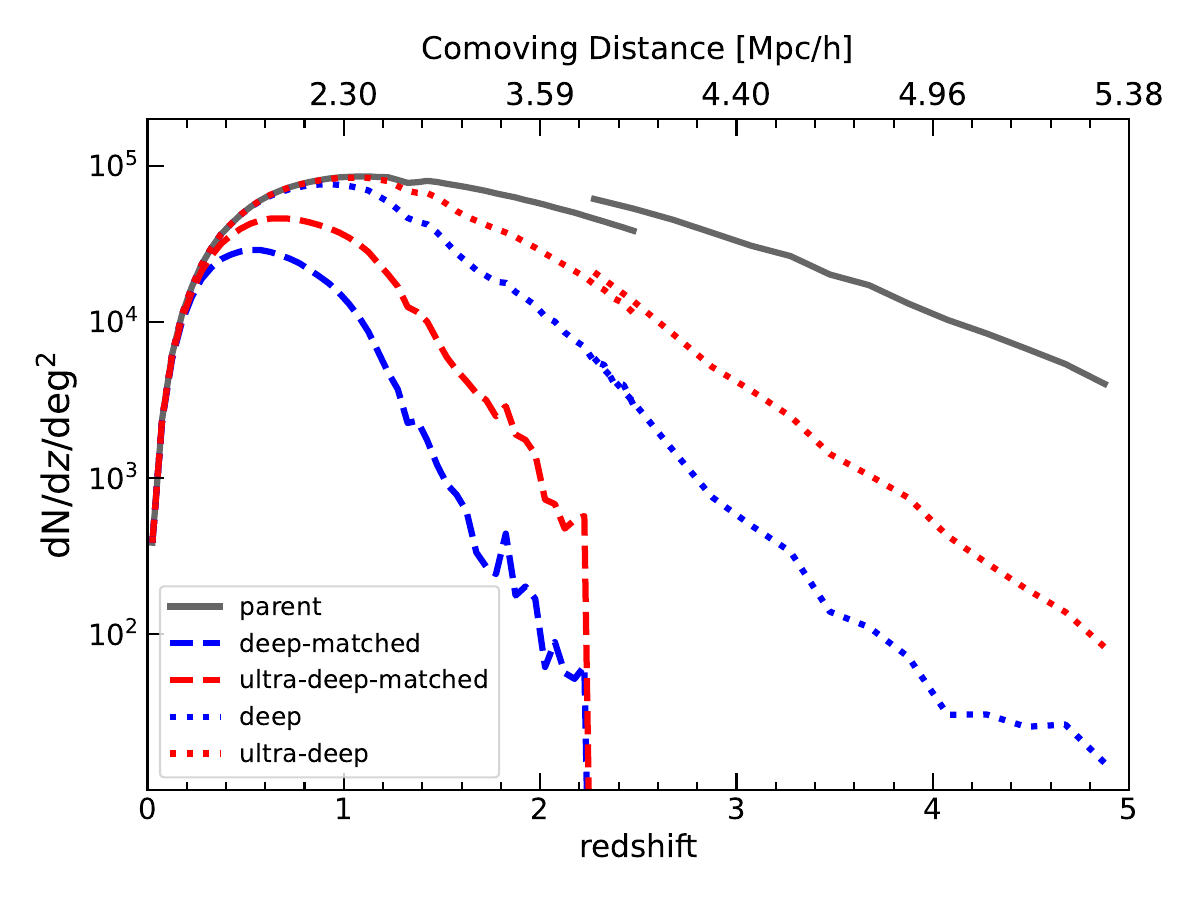}
    \includegraphics[width=0.48\linewidth]{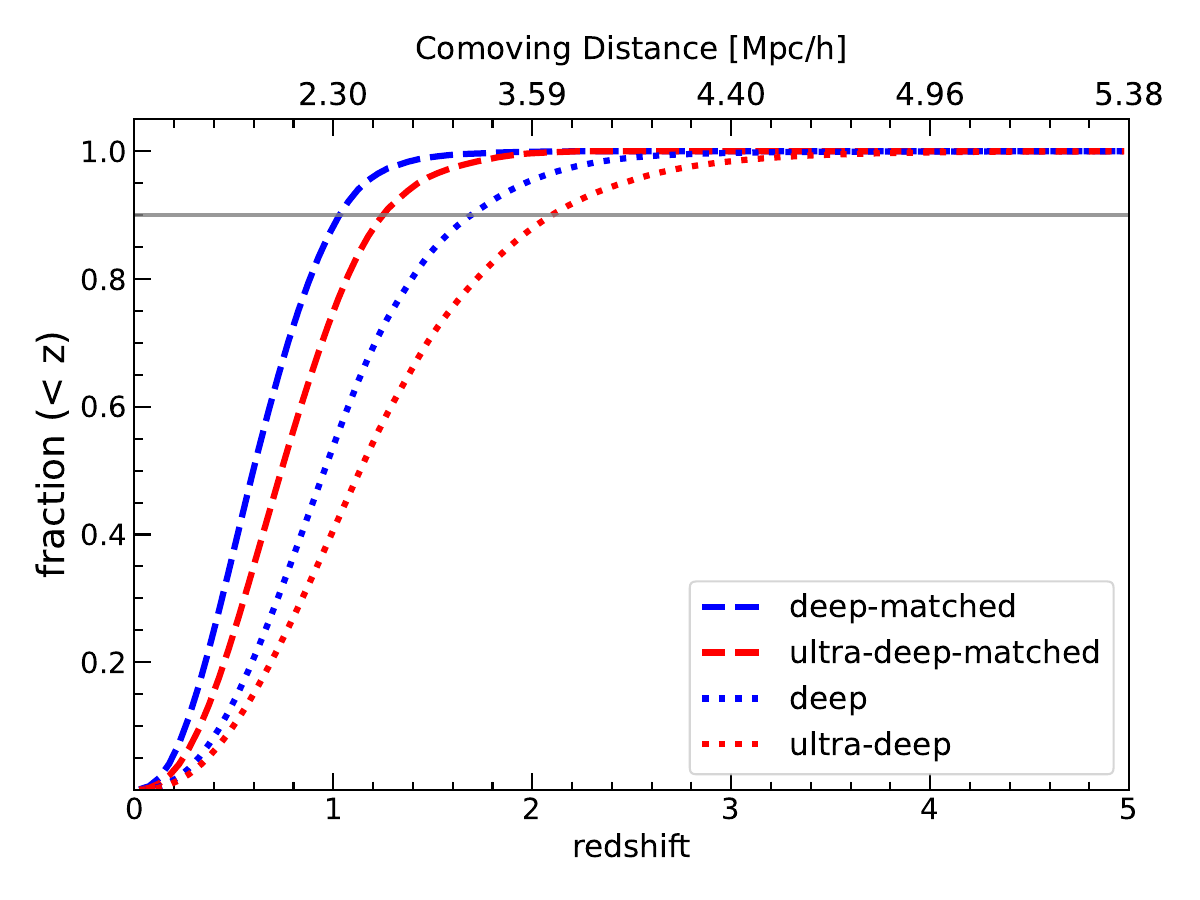}
    \caption{Redshift distributions of CSST-selected galaxies from Jiutian light-cone mocks. \text{Left}: number densities as functions of redshift. Right: cumulative redshift distributions. 
    The black lines show the stellar-mass-limited parent samples ($M_\star > 10^{8.74}~\mathrm{M_\odot}$). 
    The blue and red lines show results from the deep and ultra-deep survey selections, respectively.
    The dotted lines represent the distributions of galaxies that meet the CSST flux limit in at least one filter (any filter), while the dashed lines show those that meet the flux limits in all filters simultaneously. The gray line in the right panel marks the 90\% completeness in the cumulative distribution.}
    \label{fig:z_distri}
\end{figure*}

In Figure~\ref{fig:proj_corr} we compare the projected correlation functions of our mock galaxies with measurements from SDSS by~\citet{Li2006}. For a fair comparison with the observation, we select galaxies from a 10000 $\rm deg^2$ region in the Jiutian-1G light-cone with $r$-band apparent magnitude $14.5<\mathrm{mag}_r<17.77$ and absolute magnitude $-23<M_{0.1r}<-16$, and restrict the redshift range to $0.02 \leq z \leq 0.3$, following~\citet{Li2006}.

The projected correlation function was computed using the widely adopted Landy-Szalay estimator \citep{Landy1993ApJ},
\begin{equation}
\xi(r_p, \pi)=\frac{{DD}(r_p, \pi)-2{DR}(r_p, \pi)+{RR}(r_p, \pi)}{{RR}(r_p, \pi)}, 
\label{eqn:estimator}
\end{equation}
where $DD$, $DR$ and $RR$ are the counts of galaxy-galaxy, galaxy-random, and random-random pairs, respectively. These pairs are counted within separations $r_p$(perpendicular to the line of sight) and $\pi$ (parallel to the line of sight).
We then derive the projected correlation function $w_p(r_p)$ by integrating $\xi(r_p, \pi)$ along the line-of-sight direction,
\begin{equation}
w_p(r_p) = 2 \int_0^\infty \xi(r_p, \pi) d\pi = 2 \sum_i \xi(r_p, \pi_i) \Delta \pi_i.
\label{eqn:wprp}
\end{equation}
The summation runs from $\pi_{\rm min}=0\ \mpch$ to $\pi_{\rm max}=60\ \mpch$, with a bin size of $\Delta\pi=1\ \mpch$. For the transverse separation, we employed 20 logarithmically spaced bins covering $0.1<r_p<30\ \mpch$. 

As shown in Figure~\ref{fig:proj_corr}, the clustering of our mock galaxies shows generally good agreement with observational measurements in SDSS across the different stellar mass bins. The agreement is worse at the low mass end. This may indicate that the satellites in our model are slightly fainter than their observational counterparts. We have verified that the observational flux limit places a strong selection of the galaxy population in the lowest mass bins, and the clustering of the mock can be boosted by relaxing the flux limit of the satellite galaxies which are more clustered on small scales. 
For $10^{11}<M_\star/ \rm M_\odot<10^{11.5}$, the model overpredicts the clustering at small separations. This can be interpreted as an excess of satellite galaxies in this mass range, which could reflect insufficient merger rate of the satellites by dynamical friction or insufficient tidal mass loss of the stellar component in the model. We leave a detailed investigation of these issues and improvements over them to future work.

\section{Galaxy Distribution in the CSST Surveys}
\label{sec:forecasts}

The CSST surveys consist of a deep survey covering 17,500 deg$^2$ and an ultra-deep survey covering 400 deg$^2$ (see Figure 4 in \cite{CSSTintro} for the footprint). Both surveys will carry out simultaneous multi-band photometric and slitless spectroscopic observations. 
The expected flux limits of the photometric surveys are as follows:
\begin{itemize}
    \item Deep Survey: $g<26$, $y<24.4$, $u, r, i, z, \mathrm{NUV}<25.5$,
 \item Ultra-Deep Survey: $g<27$, $y<25.7$, $u, r, i, z, \mathrm{NUV}<26.5$,
\end{itemize}
while the spectroscopic surveys will reach $GU<22$ (23), and $GV, GI<23$ (24) for the deep (ultra-deep) surveys in three slitless spectroscopic bands. For more detailed specifications of the surveys, we direct readers to the CSST introduction paper~\cite{CSSTintro}.

As the spectroscopic surveys are shallower, in the following we only construct light-cone mock galaxy catalogs for CSST following the photometric flux limits. All the results below are based on galaxies from light-cone catalogs. We construct a full-sky light-cone catalog for CSST extending to $z=5$, combining outputs from Jiutian-1G ($0<z<2.5$) and Jiutian-2G ($2.2<z<5$). This hybrid approach leverages the higher resolution of Jiutian-1G for low-redshift faint objects and the larger volume of Jiutian-2G for high-redshift statistics. The redshift overlap ($2.2<z<2.5$) enables consistency verification between the simulations. We release the light-cone catalogs from the two simulations separately.

\begin{figure}[H]
    \includegraphics[width=0.48\textwidth]{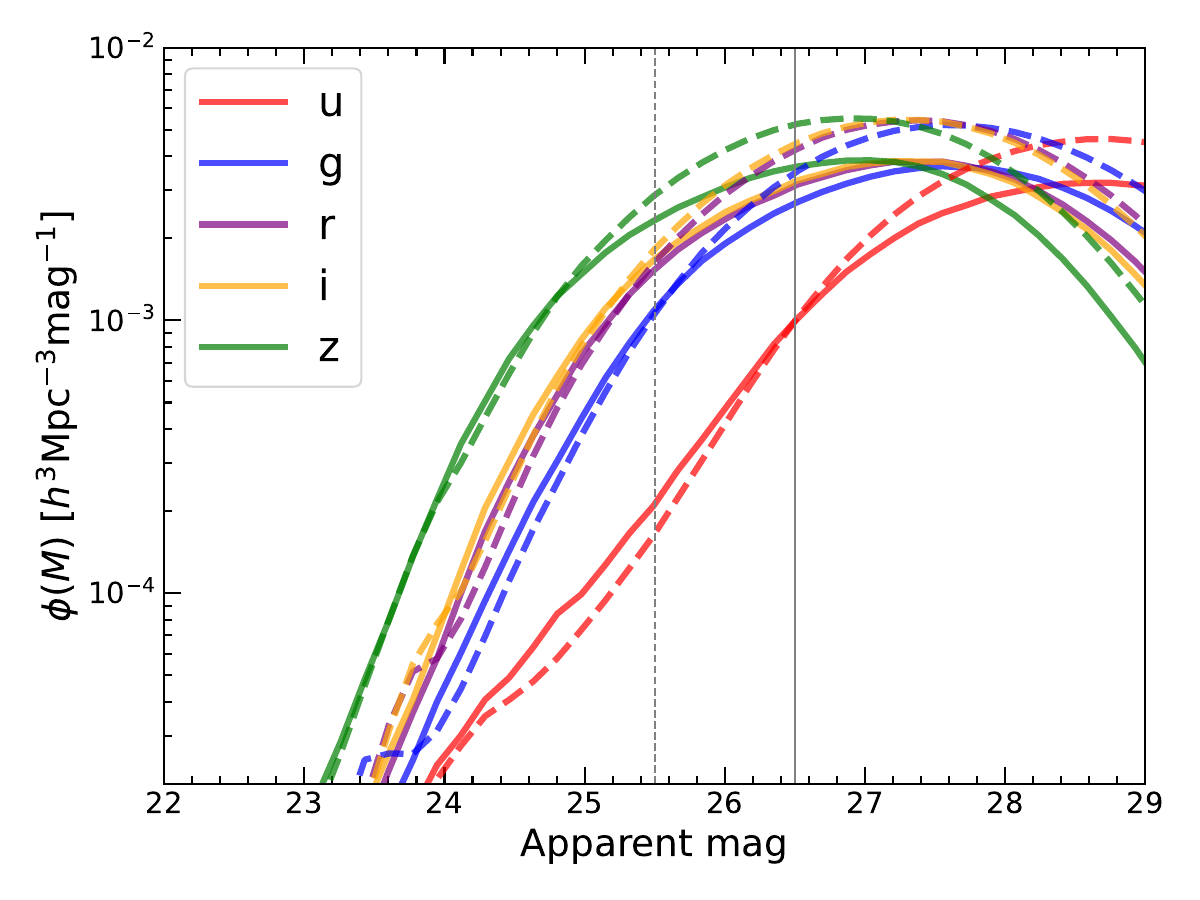}
    \caption{Luminosity function comparison at $z=2.2$, the connecting redshift in Figure~\ref{fig:z_distri}, between Jiutian-1G (solid lines) and Jiutian-2G (dashed lines). The different colors distinguish different CSST filters. The dashed and solid vertical lines represent the deep and ultra-deep flux limits in $u,r,i,z$ bands. The flux limit of $g$ band is 0.5 mag fainter than vertical lines shown in the figure. }
    \label{fig:LF_comp}
\end{figure}

\begin{figure*}[htbp]
    \centering
    \includegraphics[width=0.99\linewidth]{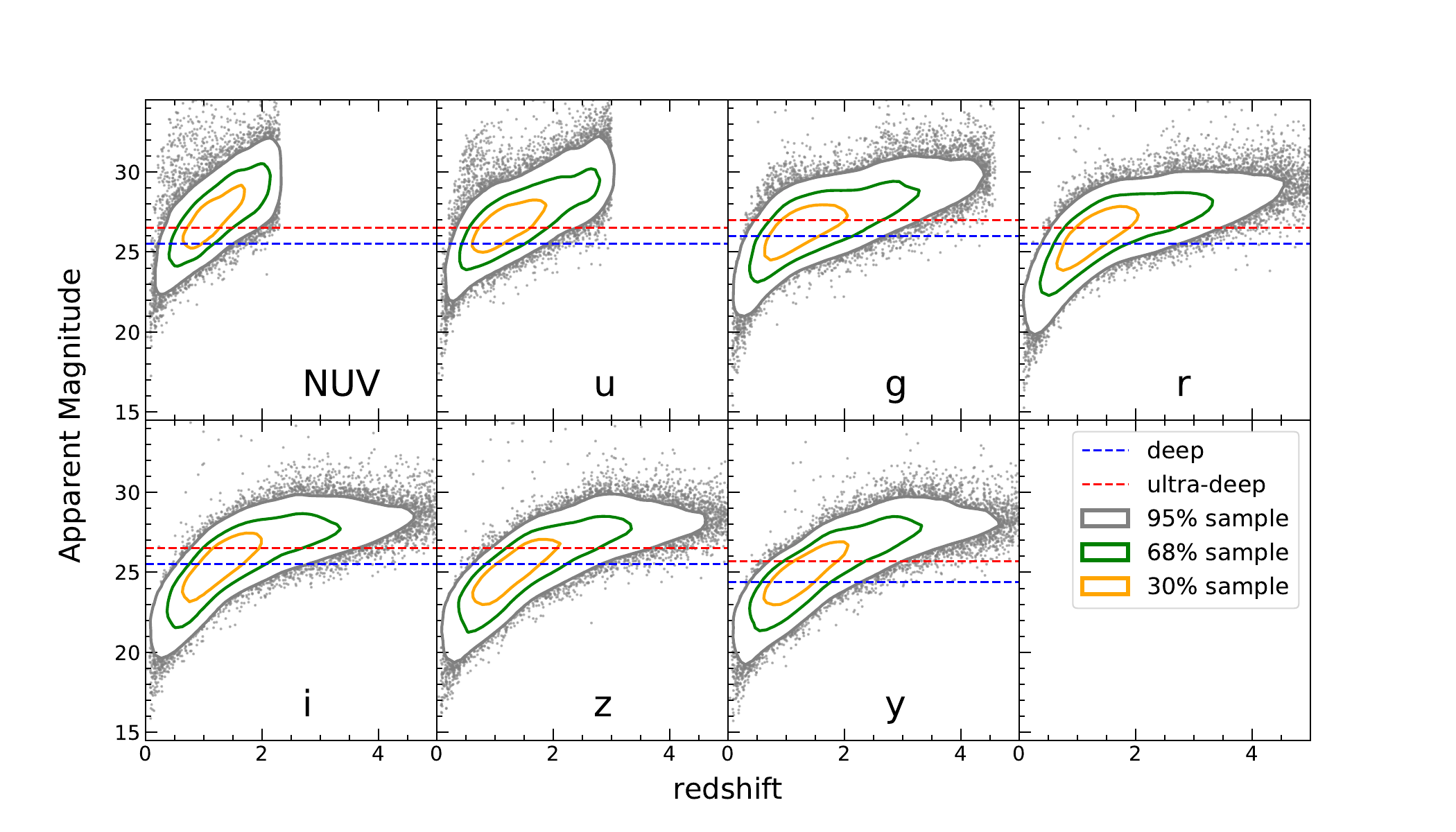}
    \caption{Apparent magnitude-redshift distributions of mock galaxies in the mock CSST light-cone. The red and blue dashed lines show the flux limits for the deep and ultra-deep surveys, respectively. The coloured contours mark the 95\%, 68\%, and 30\% most populated regions of the distribution.}
    \label{fig:mag_distri}
\end{figure*}

Figure~\ref{fig:z_distri} reveals several important characteristics of the CSST survey by showing the galaxy redshift distributions in our light cone samples using the CSST flux limits mentioned before. The distributions of stellar-mass limited samples (black solid lines in left panel) show discrepancies between Jiutian-1G and 2G at the transition region, reflecting the slightly different stellar mass functions between the two runs at $z\sim 2.2$ as shown in Figure~\ref{fig:sm}.
However, this discontinuity has little influence on the flux-limited samples. As shown by the dotted lines, once the flux limits are applied, the predictions from the two simulations join smoothly around the intersecting redshift ($z\sim2.2$). To understand this behavior in more detail, in Figure~\ref{fig:LF_comp} we show the luminosity functions from the two simulations around the intersecting redshift. In each band, the luminosity function from Jiutian-2G lies above the corresponding luminosity function from the Jiutian-1G only at the faint end. Once the CSST flux limits are applied, the difference between the two diminishes.

If we further require galaxies to be observable across multiple CSST bands, the redshift distribution truncates sharply at $z\approx 2.2$. This is mainly because the Lyman-break shifts to the NUV band around this redshift, making the galaxies inaccessible in NUV. Figure~\ref{fig:mag_distri} shows the influence of the CSST flux limits in different bands more clearly. It can be observed that the NUV band places the tightest limit on the redshift distribution. The NUV band exhibits a pronounced drop at $z \approx 2.2$, followed by the $u$ band at $z \approx 2.8$ and the $g$ band at $z \approx 4.5$, due to the redshifted Lyman break transiting these filter bandpasses. 
At higher redshifts beyond $z>3$, the $r,\ i,\ z$ band samples become increasingly dominant as the optical emission redshifts into these longer wavelength filters.

In the right panel of Figure~\ref{fig:z_distri}, we present the redshift cumulative number fractions for all four different survey conditions, combining the light-cone galaxies at $z<2.5$ from Jiutian-1G and those at $z>2.5$ from Jiutian-2G. It shows that about 90\% of the galaxies from the deep survey with joint selections reside at redshifts below $z=1$, while the ultra-deep survey extends this depth to $z=1.2$. If we do not require the galaxies to be jointly observable across bands, the limits become $z=1.7$ and $z=2.1$.

\begin{figure*}[htbp]
\centering
    \includegraphics[width=0.49\linewidth]{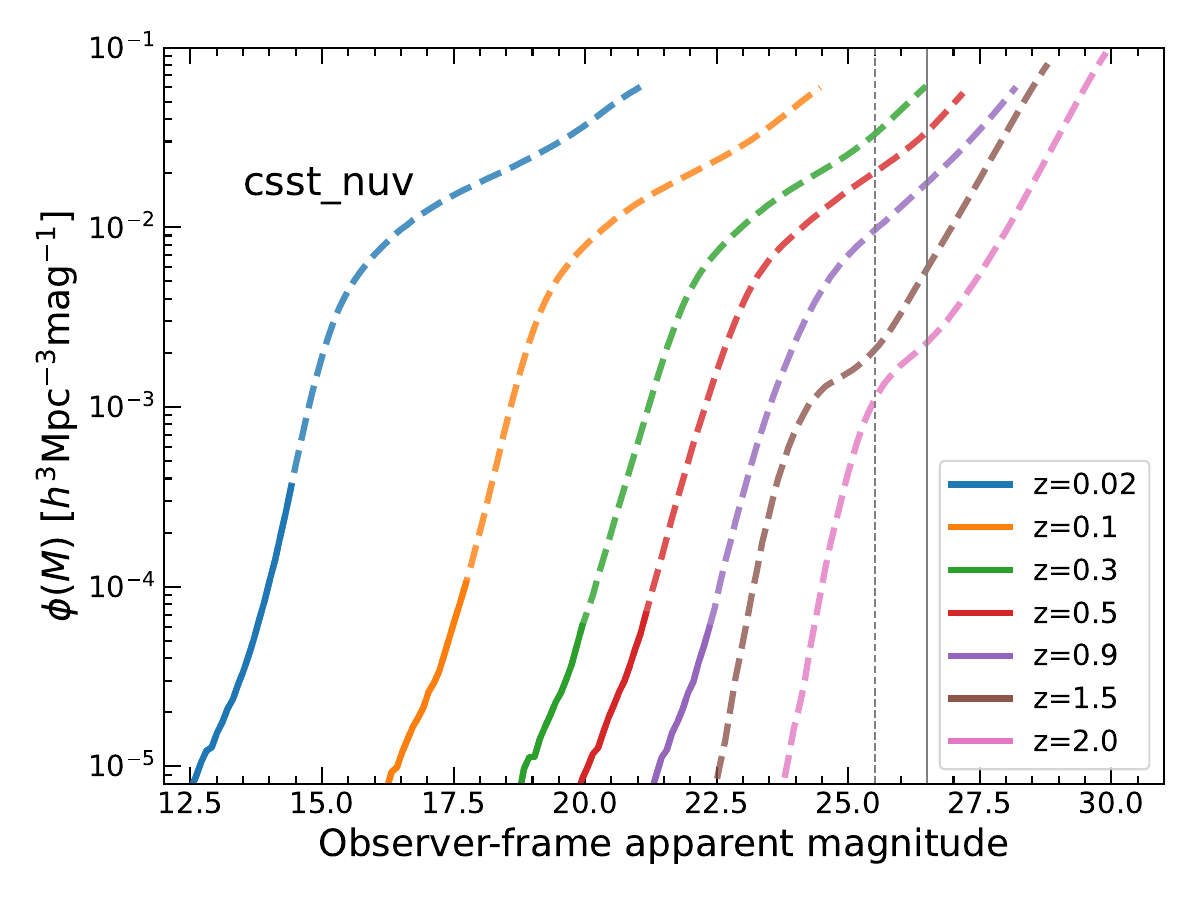}
    \includegraphics[width=0.49\linewidth]{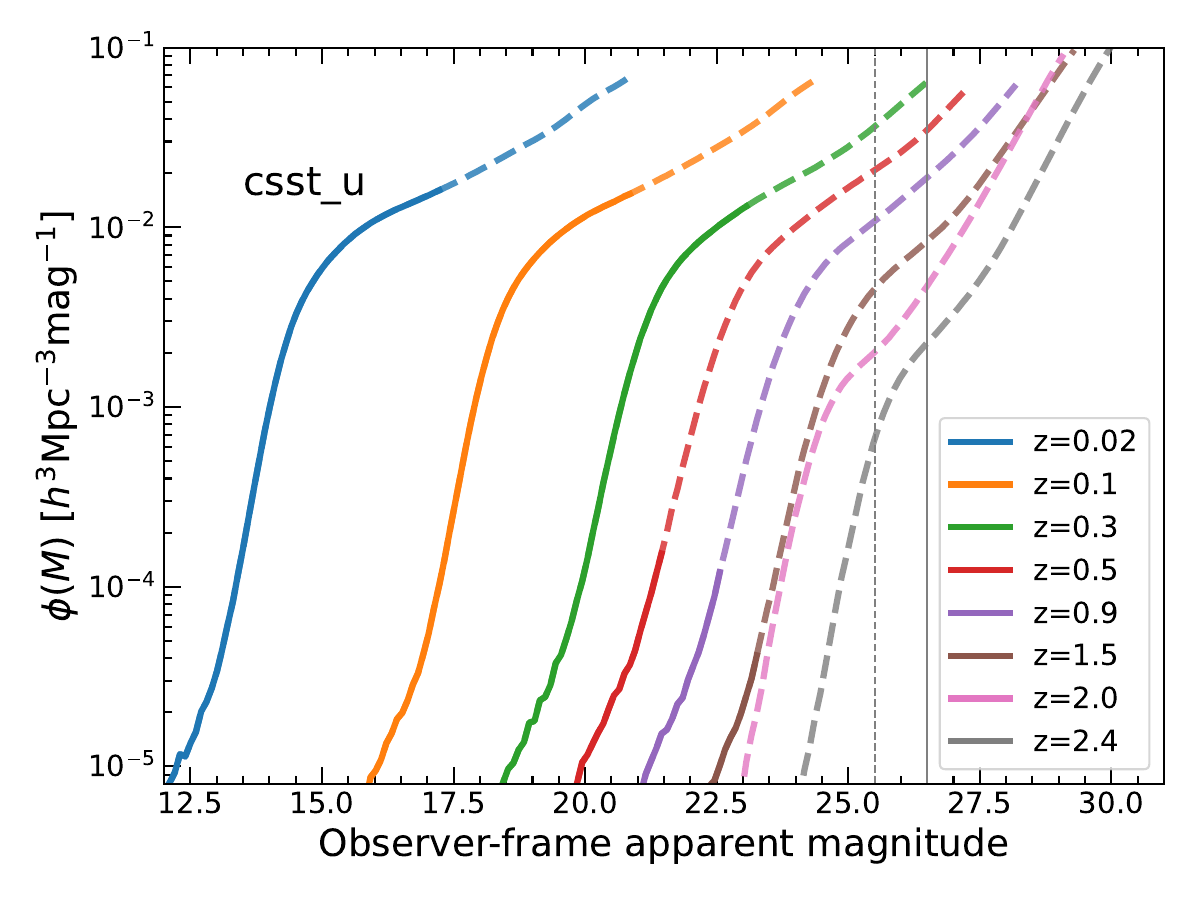}
    \caption{Observer-frame luminosity function evolution in the CSST NUV (left) and $u$ (right) bands. The solid segment of each curve indicate galaxies brighter than the soft magnitude limits, while dashed segments correspond to galaxies with magnitudes between the soft and hard limits. Vertical gray lines mark the flux limits for the deep (dashed) and ultra-deep (solid) surveys.}
    \label{fig:csst_predi}
\end{figure*}

One of the major advantages of CSST compared with previous generations of major optical surveys is the coverage up to the NUV band, which is a sensitive probe of the formation of massive stars and plays an important role for the estimation of star formation rate and photometric redshift.
Figure~\ref{fig:csst_predi} predicts the evolution of NUV and $u$-band luminosity functions as a function of observer-frame magnitudes at different redshifts. 
The CSST ultra-deep survey will be able to probe the faint end of the luminosity function up to $z\sim 1.5$ in the NUV band, and to $z\sim 2$ in the $u$ band.
This exquisite depth is comparable to that achieved by UVCANDELS \citep{wangLymanContinuumEscape2025,sunUltravioletLuminosityFunction2024}, which reaches a 5-$\sigma$ limiting magnitude of AB $\approx$ 27 for compact sources; despite being the largest extragalactic HST program, it covers only $\sim$426 arcmin$^2$ over HST’s 30-year nominal mission lifetime.

\section{Summary and Conclusions}
\label{sec:conclu}
Based on the comprehensive pipeline developed in this work, we present a light-cone tailored for the China Space Station Telescope (CSST). Starting from the Jiutian suite of large-volume, high-resolution $N$-body simulations, halo and subhalo merger trees are constructed with the time domain subhalo finder \hbt. The \gaea semi-analytical model of galaxy formation has been adapted to run over the \hbt merger trees in order to generate mock galaxy catalogues. Starting from their predicted SFHs, we adapt a neural network-based synthesizer, \starduster, to compute the spectral energy distribution and broadband magnitudes of each galaxy, incorporating a detailed treatment of interstellar dust attenuation calibrated using radiative transfer simulations.
We develop a dedicated light-cone builder code (\blic) which interpolates galaxy properties from discrete snapshots onto a continuous light-cone. 
We test the performance of different interpolation schemes, and we find that cubic interpolation provides the optimal scheme to reconstruct galaxy trajectories.

We apply the pipeline to construct light-cone mocks for CSST up to redshift $z=5$, combining outputs from two simulations, Jiutian-1G covering $0<z<2.5$ and Jiutian-2G covering $2.2<z<5$. The statistical properties of galaxies show generally good convergence across the two runs, and the two light-cone catalogs connect smoothly at their overlapping redshift after the survey selection cut.

Validation against extensive observational datasets confirms that our model successfully reproduces key galaxy population statistics, including stellar mass functions up to $z\approx1$, gas fraction scaling relations over a wide range in stellar mass, size-mass relations up to redshift $\sim2$, and projected correlation functions measured from SDSS. Notable exceptions appear in the stellar mass functions at $z>1$ where the model under-predicts massive galaxies, reflecting a deficit in the high-redshift star formation rate in the adopted version of the \gaea model. 

Forecasts for the CSST surveys demonstrate the telescope's exceptional capability to probe galaxy evolution across cosmic epochs. The deep survey (17,500 $\rm deg^2$) will detect a high density of $\sim 20/{\rm arcmin}^2$ galaxies up to $z\approx1.5$ (90\% upper limit) in all seven bands while observations in $g,r,i,z$ bands alone allow for deeper redshift limits up to $z\approx 2.1$ with a number density of $\sim 31/{\rm arcmin}^2$. The number densities of the ultra-deep survey will be 1.5 times larger than that of the deep survey. 

Characteristic Lyman-break signatures produce sharp truncations in the redshift distributions, particularly evident in the NUV band at $z\approx2.2$ and in the $u$ band at $z\approx 2.8$. The faint end of the luminosity functions in the NUV and $u$-bands can be probed up to $z\sim 2$. 

The mock catalogs constructed, as well as the corresponding pipeline developed, represent a significant effort towards the science preparation and facility building for the CSST survey. Future work will build upon and extend these efforts to further improve our physical understanding of galaxy formation and quantitative modelling of survey systematics, to exploit the full potential of CSST in studying galaxy formation and cosmology. The mock catalogs produced here will be publicly available on the Jiutian website at \url{https://jiutian.sjtu.edu.cn}.

We also want to highlight some caveats in the current version of the mock catalogs presented here. We have already illustrated that the high redshift star formation rate differs from observational measurements, and the magnitude distribution of satellite galaxies could differ from observations. Related to this, the observational color distribution of galaxies is notoriously difficult to reproduce from theoretical models including ours~\citep[e.g.,][]{2013MNRAS.429..556M, Henriques2015MNRAS, Trayford2015MNRAS.452.2879T, 2017MNRAS.471.1671D, Donnari2019MNRAS.485.4817D, Akins2022ApJ...929...94A, XuXiaoju2022MNRAS.516.4276X}. Besides, we have not accounted for some observational data reduction procedures such as source detection and sky background subtraction which may introduce further systematics into the photometry.  
Users are encouraged to perform their own assessment on the validity of the mock catalogs for specific science cases, especially when such comparisons are not covered in this paper.

{\bf Acknowledgements:}

This work is supported by NSFC(12573022,12273021), National Key R\&D Program of China (2023YFA1607800, 2023YFA1607801, 2023YFA1605600, 2023YFA1605601), science research grants from the China Manned Space Project (No. CMS-CSST-2021-A03, CMS-CSST-2021-A02), and 111 project (No. B20019). We also acknowledge the Yangyang Development Fund. The computation of this work is done on the Gravity supercomputer at the Department of Astronomy, Shanghai Jiao Tong University. This research has made use of the SVO Filter Profile Service "Carlos Rodrigo"\cite{2020sea..confE.182R, Rodrigo2024A&A...689A..93R}, funded by MCIN/AEI/10.13039/501100011033/ through grant PID2023-146210NB-I00. 
YSQ acknowledges support from Key R\&D Program of Zhejiang, China (Grant No. 2024SSYS0012). XW is supported by the China Manned Space Program with grant no. CMS-CSST-2025-A06, the National Natural Science Foundation of China (grant 12373009), the CAS Project for Young Scientists in Basic Research Grant No. YSBR-062, the Fundamental Research Funds for the Central Universities, and the Xiaomi Young Talents Program. 

Conflict of Interest  The authors declare that they have no conflict of interest.

\bibliography{sample631}{}
\bibliographystyle{aasjournal}

\newpage
\section*{Supplementary Materials}

\section{Interpolation Scheme Comparisons}
\label{sec:interpolation}

\subsection{Interpolation Schemes}

As a crucial step in building a light-cone, the interpolation of trajectories directly determines the accuracy of the solved coordinates of objects in the light-cone, which subsequently affects the accuracy of other interpolated properties. Below we describe and compare five interpolation schemes, some of which have been adopted in previous works~\citep[e.g.,][]{2013MNRAS.429..556M, 2022MNRAS.516.4529S}.

\subsubsection{Linear interpolation}

The linear interpolation connects two positions at time $t_{i}$ and $t_{i+1}$ with the linear equation: 
\begin{equation}
    \vec{X}\left(t\right)=\vec{A}t+\vec{B}.
\end{equation}
The parameters $\vec{A}$ and $\vec{B}$ are determined from the boundary conditions $\vec{X}(t_i)=\vec{A}t_{i}+\vec{B}$ and $\vec{X}(t_{i+1})=\vec{A}t_{i+1}+\vec{B}$. The interpolated velocity is simply 
\begin{equation}
    \vec{v}(t)=\vec{A}. 
\end{equation}

\subsubsection{Cubic interpolation}

The cubic interpolation extends linear interpolation by taking velocities into consideration. At time $t_{i}$ and $t_{i+1}$, the cubic interpolation is written as
\begin{align}
    \vec{X}(t)&=\vec{A}t^3+\vec{B}t^2+\vec{C}t+\vec{D},\\
    \vec{v}(t)&=3\vec{A}t^2+2\vec{B}t+\vec{C}.
\end{align}
The four parameters are determined by four boundary conditions at two different snapshots: 
\begin{align}
    \vec{X}\left(t_i\right)&=\vec{A}t_i^3+\vec{B}t_i^2+\vec{C}t_i+\vec{D},\\\vec{X}\left(t_{i+1}\right)&=\vec{A}t_{i+1}^3+\vec{B}t_{i+1}^2+\vec{C}t_{i+1}+\vec{D},\\
    \vec{v}\left(t_i\right)&=3\vec{A}t_i^2+2\vec{B}t_i+\vec{C},\\
    \vec{v}\left(t_{i+1}\right)&=3\vec{A}t_{i+1}^2+2\vec{B}t_{i+1}+\vec{C}.
\end{align}

\subsubsection{Interpolations in polar and spherical coordinates}

This approach, inspired by~\citet{2013MNRAS.429..556M} with modifications, constrains polar-interpolated satellites to an orbital plane. The plane is defined by the positions of the satellite at two subsequent snapshots and that of the central object. Between time $t_{i}$ and $t_{i+1}$, the satellite orbit is interpolated as
\begin{equation}
    r\left(t\right)=At+B,
\end{equation}
\begin{equation}
    \theta\left(t\right)=Ct+D,
\end{equation}
where $r$ and $\theta$ are the polar coordinates of the satellite relative to the central object. The four parameters can be fixed from the relative coordinates of the satellite at the bracketing snapshots, eliminating velocity requirements.

The spherical interpolation operates similarly without planar constraints. Here, we convert the relative position of the satellite from Cartesian to spherical coordinates, and then interpolate using only positional data. 

\subsubsection{Cubic spline interpolation}

The cubic spline is composed of a piecewise cubic polynomial that is twice continuously differentiable. We implement this method using the \textsc{CubicSpline} function from \textsc{scipy}\cite{2020SciPy-NMeth} to interpolate the trajectory of objects.
A meaningful application of the cubic spline requires at least three snapshots, while more snapshots provide more constraints on the global shape of the trajectory. In the tests, we utilize positions from 9 preceding and 9 subsequent snapshots relative to the target time when available.

\subsection{Tests}

To benchmark the five interpolation schemes, we conduct two tests using the Illustris-3 simulation snapshots~\cite{Vogelsberger2014Natur, Genel2014MNRAS, Vogelsberger2014MNRAS} with a box size of $L=75\,\mathrm{Mpc}/h$ and cosmological parameters, $\Omega_{M}=0.2726$, $\Omega_\Lambda=0.7274$, $H_{0}=70.4\,\mathrm{km}/\mathrm{s}/\mathrm{Mpc}$. This small volume reduces the computational cost.
We select two snapshot triplets with differing time intervals, with Set I covering snapshots 67 to 69 with a time separation of  $\Delta t_\mathrm{I}=0.324\,\mathrm{Gyr}$ between the first and third snapshots, and Set II covering snapshots 50-52 with $\Delta t_\mathrm{II}=0.176\,\mathrm{Gyr}$.
For each set, we interpolate between the first and third snapshots to predict the positions at the intermediate snapshot, and then compare with the ground-truth data. 

\subsubsection{Convergence}

\begin{figure}[H]
    \includegraphics[width=0.48\textwidth]{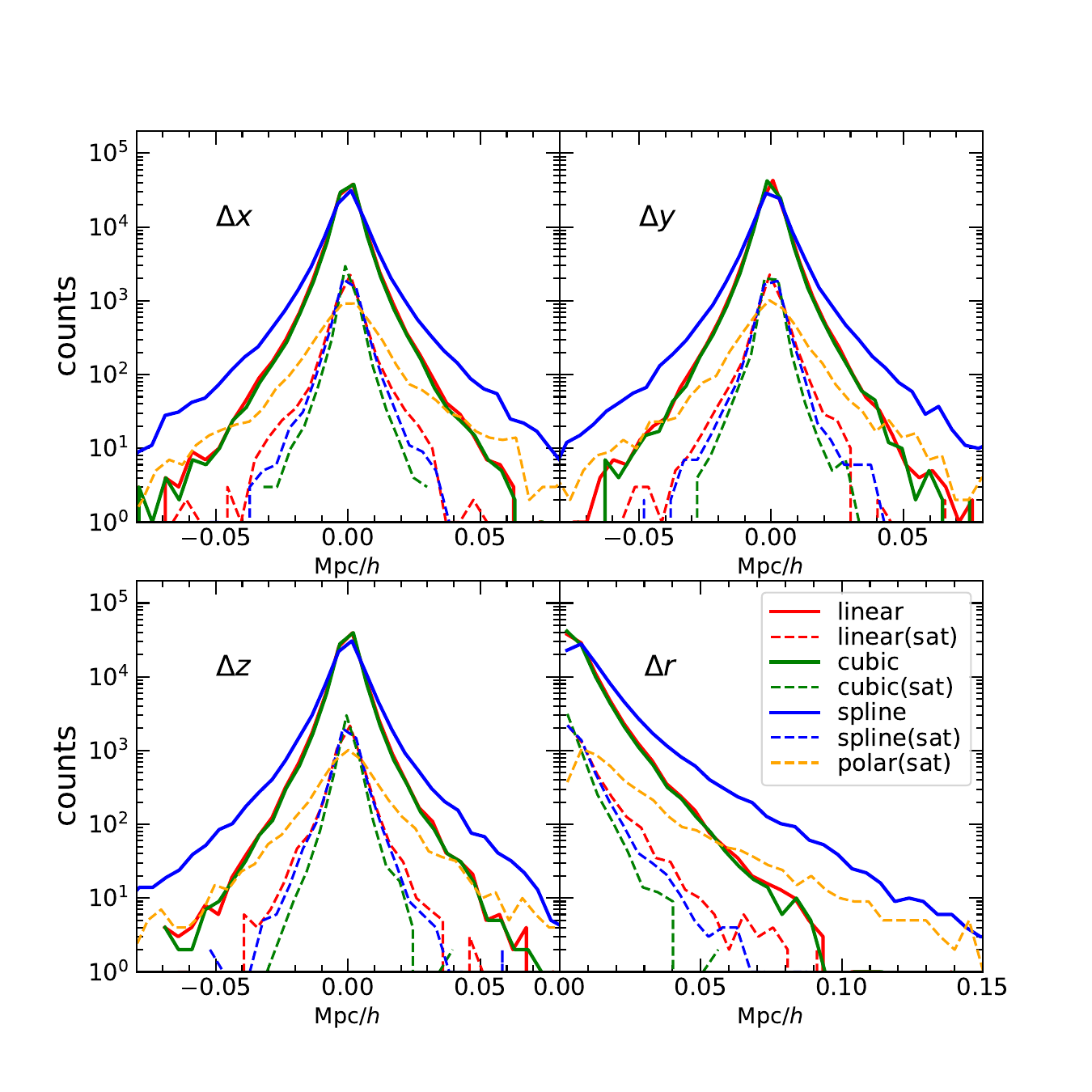}
    \caption{
    Same as Figure 2 in the main text, positional deviations ($|\Delta X| = |X_\mathrm{pred}-X_\mathrm{true}|$) between interpolated and actual snapshots. Solid lines: all objects; dashed lines: satellites only (polar interpolation exclusively applies to satellites). Panels show $\Delta x$, $\Delta y$, $\Delta z$, and $\Delta r$ (where $\Delta r = \sqrt{\sum \Delta x_i^2}$).
    Prediction for snapshot 51 ($\Delta t = 0.176\,\mathrm{Gyr}$; 87,759 objects, 4,475 satellites). As the time interval is shorter, linear and cubic interpolations outperform cubic spline for total samples. Polar interpolation underperforms other schemes for satellites. }
    \label{fig:pos_comp}
\end{figure}

We quantify the positional accuracy of the interpolations through the component-wise deviations ($\Delta x, \Delta y, \Delta z$) and total displacement $\Delta r=\sqrt{\Delta x^2+\Delta y^2+\Delta z^2}$. The results for Set I are already presented in the main text, while Figure~\ref{fig:pos_comp} shows the results for Set II. 
Narrower distributions centered at zero indicate higher accuracy. These results show that the cubic interpolation consistently outperforms others across time intervals and object types. The cubic spline improves relative to linear interpolation at larger time intervals ($\Delta t = 0.324\,\mathrm{Gyr}$), while polar interpolation exhibits the largest errors.

\subsubsection{Correlation function}

\begin{figure*}[!htbp]
\centering
    \includegraphics[width=0.48\textwidth]{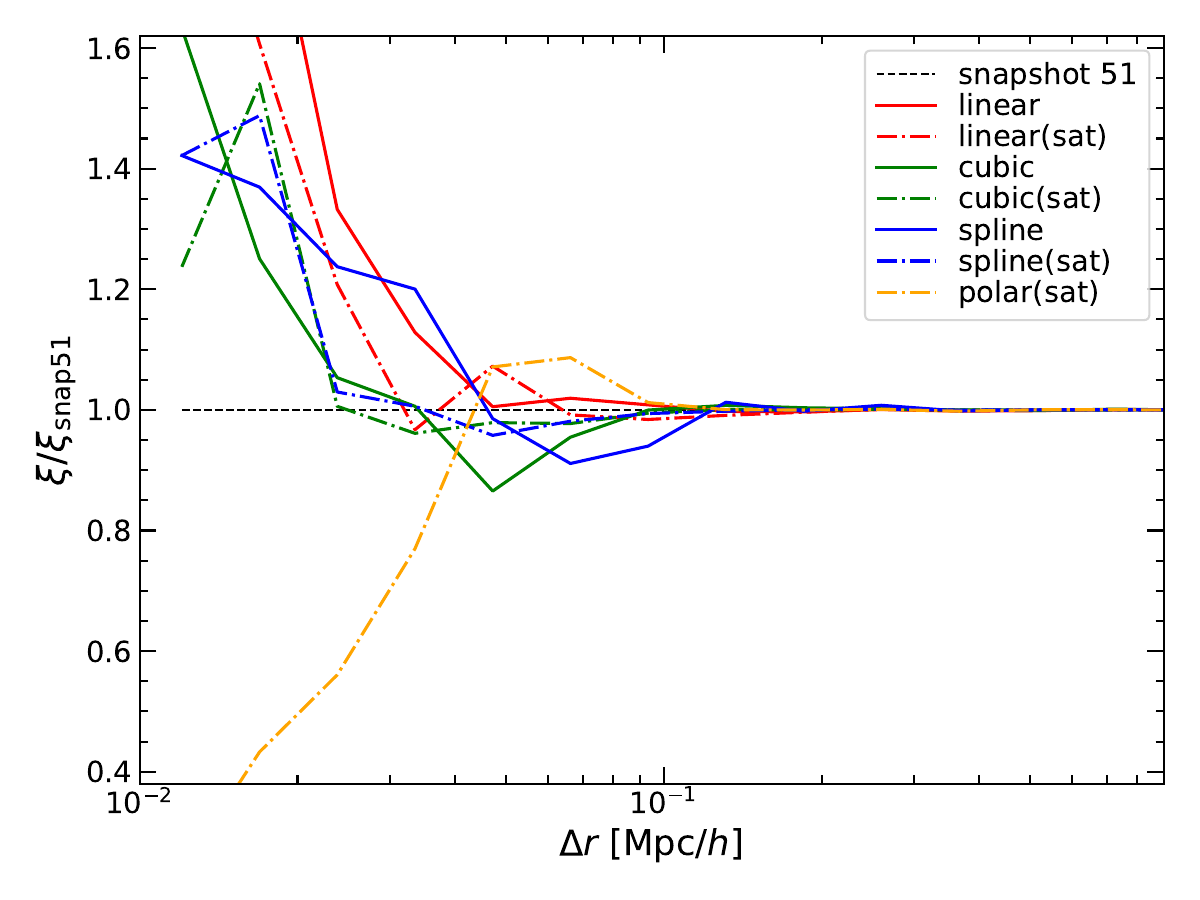}
    \includegraphics[width=0.48\textwidth]{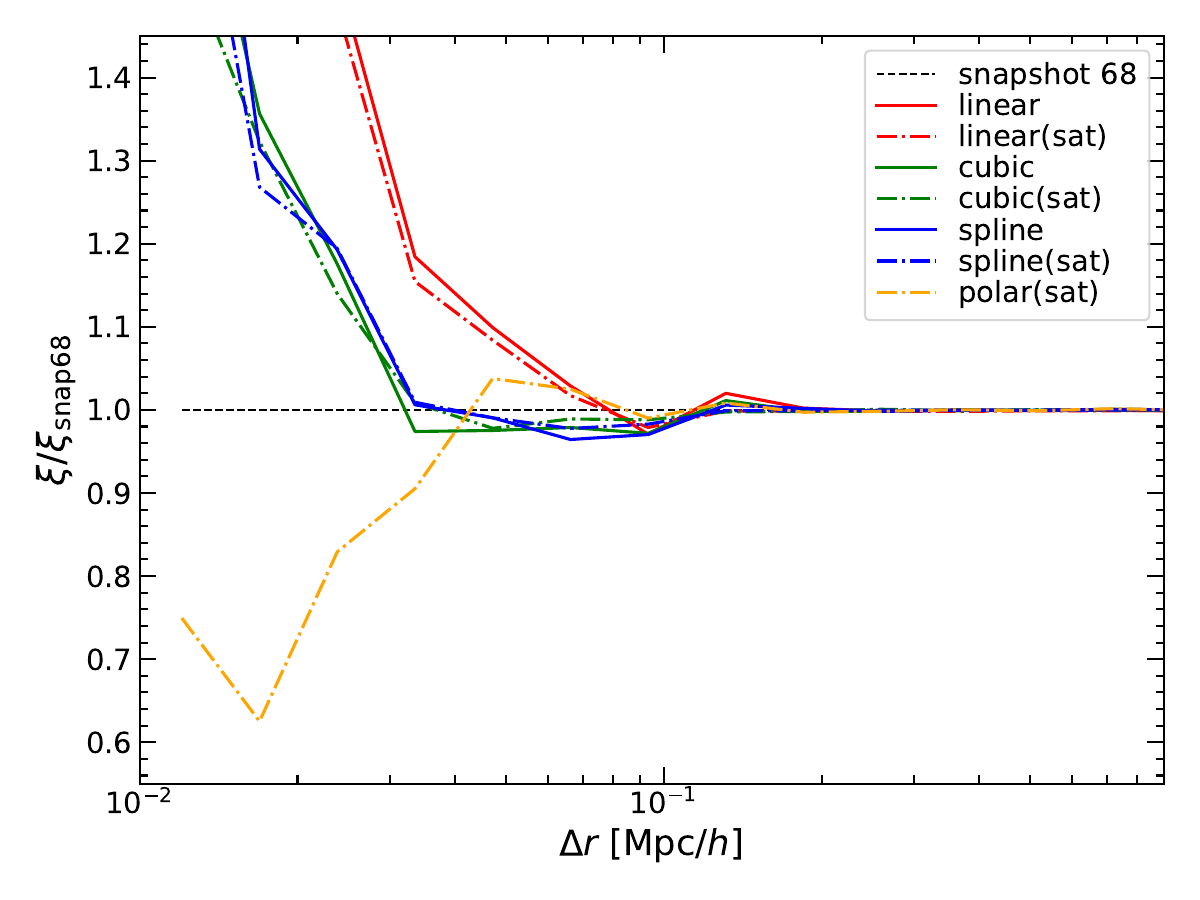}
    \caption{
    Two-point correlation function (2PCF) ratios $\xi_{\mathrm{pred}}/\xi_{\mathrm{true}}$. Solid lines: all objects; dashed lines: predicted satellites paired with true centrals. The black dashed line indicates perfect agreement ($\xi_{\mathrm{pred}}/\xi_{\mathrm{true}} = 1$). \textbf{Left}: Snapshot 51 ($\Delta t = 0.176\,\mathrm{Gyr}$). \textbf{Right}: Snapshot 68 ($\Delta t = 0.324\,\mathrm{Gyr}$).}
    \label{fig:corr}
\end{figure*}

We further assess structural preservation using the two-point correlation function (2PCF) calculated via the Landy-Szalay estimator~\citep{Landy1993ApJ}.
Figure~\ref{fig:corr} shows the ratio, $\xi_{\mathrm{pred}}/\xi_{\mathrm{true}}$, for both test sets. The cubic interpolation shows slight advantages for the shorter time interval and comparable accuracy to the spline and polar interpolations in the larger time interval. All schemes converge to the true correlation function beyond $r \gtrsim 0.1-0.2\,\mathrm{Mpc}/h$, regardless of the time interval.

\subsubsection{Conclusion}

These analyses reveal that the cubic interpolation delivers optimal accuracy across time intervals and metrics (positional accuracy and correlation preservation). While cubic spline improves at larger intervals ($\Delta t = 0.324\,\mathrm{Gyr}$), linear interpolation exhibits significant degradation with increasing time intervals. Polar interpolation shows poor convergence despite its theoretical motivation for satellites. We therefore select cubic interpolation for our light-cone construction pipeline.

\section{Data Description}

Due to the enormous size of the total simulation data, we process and release it in subdivided volumes. The simulations are partitioned into 1000 sub-volumes based on the positions of galaxies (subhalos) at $z=0$. Each sub-volume thus represents 1/1000 of the entire simulation box. For the Jiutian-1G simulation, this corresponds to a sub-volume box size of 100 Mpc/$h$, while for Jiutian-2G, it is 200 Mpc/$h$. However, since particle motion occurs within the global simulation domain, we cannot guarantee that objects in each sub-volume at higher redshifts remain confined to their original $z=0$ sub-box boundaries. Nevertheless, we can still treat each sub-volume as representing 1/1000 of the total volume when computing statistical properties at higher redshifts.

\begin{figure}[H]
    \includegraphics[width=0.5\textwidth]{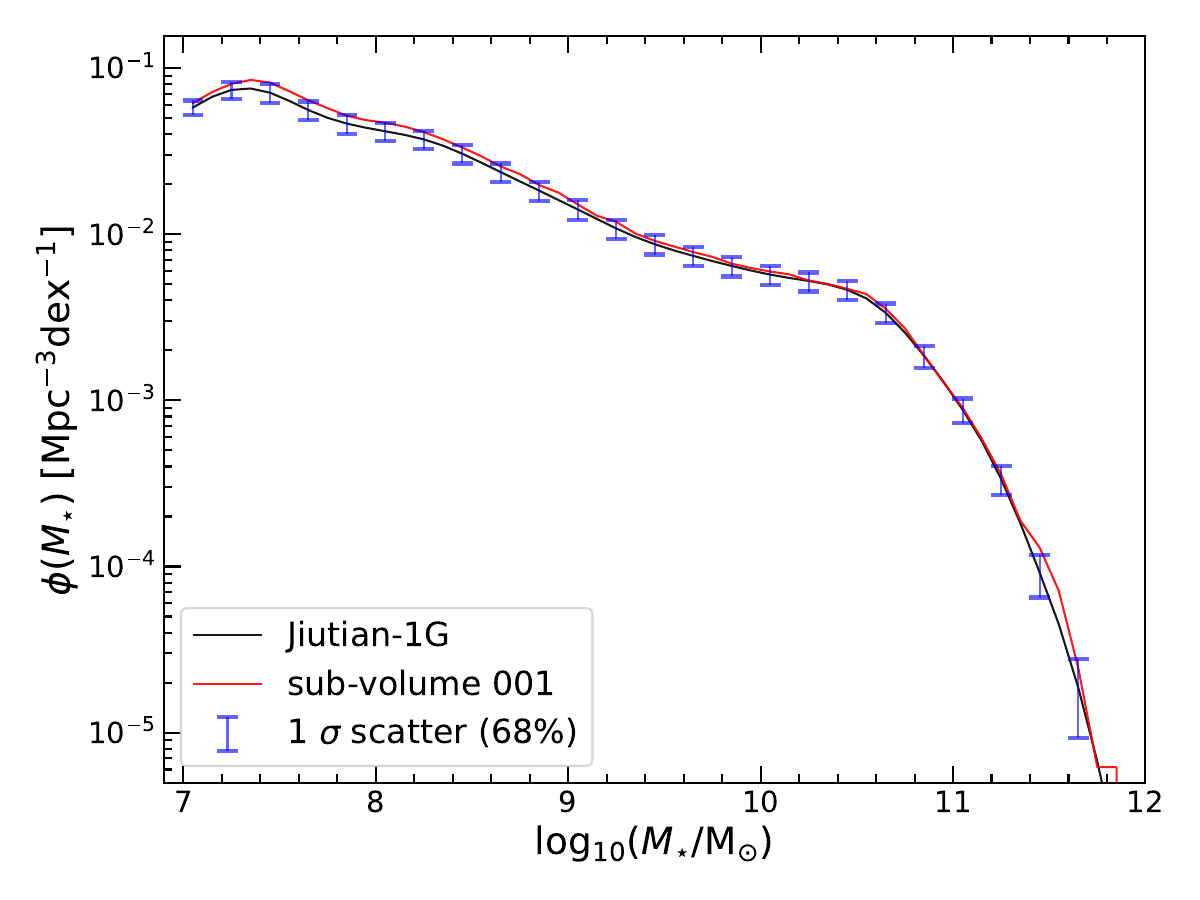}
    \caption{Galaxy stellar mass function statistics for the Jiutian-1G simulation at $z=0$. The black line represents the mean SMF computed from the entire simulation volume (equivalent to the red line in Figure 4 in the main text). The blue error bars indicate the 1$\sigma$ (68\%) scatter of the SMF distribution across sub-volumes. The solid red line shows the SMF of the released sub-volume at $z=0$.}
    \label{fig:sub_scatter}
\end{figure}

Figure~\ref{fig:sub_scatter} illustrates the statistical variance across sub-volumes in the galaxy stellar mass function at $z=0$. The black line represents the mean stellar mass function derived from the entire simulation box, corresponding to the red line shown in the left panel of Figure 4 in the main text. The blue error bars indicate the 1$\sigma$ (68\%) scatter of the SMF distribution across sub-volumes. The red line specifically denotes the SMF of the released sub-volume at $z=0$.

Part of the data described in this work are released publicly through the Jiutian website, \url{https://jiutian.sjtu.edu.cn}. 
Our first data release includes:
\begin{itemize}
    \item \gaea snapshot outputs of a subvolume, 
    \item \hbt snapshot outputs of the same subvolume, 
    \item A 5000 $\rm deg^2$ continuous CSST deep field light-cone, 
    \item Eight CSST ultra-deep field light-cones, each of $50\ \rm deg^2$.
\end{itemize}

All released files are in \texttt{hdf5} format, with the main content being an array of compound data (subhalo or galaxy) that can be easily loaded with any hdf5 libraries or utilities. The data fields for the \gaea, \hbt and light-cone files are described in Table~\ref{SAMtableA}, Table~\ref{HBTtableA} and \ref{LCtable}, respectively.

\end{multicols}

\newpage
\begin{center}
\begin{longtable}{p{0.17\textwidth}p{0.12\textwidth}p{0.7\textwidth}}
\caption{Data fields of the \gaea galaxy catalogs.}\\
\hline\hline
\multicolumn{1}{l}{Field} & \multicolumn{1}{l}{Unit} & \multicolumn{1}{l}{Description} \\

\endfirsthead

\multicolumn{3}{c}{{\tablename\ \thetable{} -- Continued}} \\
\hline\hline
\multicolumn{1}{l}{Field} & \multicolumn{1}{l}{Unit} & \multicolumn{1}{l}{Description} \\
\hline\hline
\endhead

\hline
\multicolumn{3}{r}{{Continued on next page}} \\
\endfoot

\hline\hline
\endlastfoot

\hline\hline
        GalID & - & The unique ID of the galaxy in the simulation, identical to the TrackId field in the \hbt tables. \\ \hline
        SinkGalID & - & The unique ID of the galaxy that the current galaxy will merge into. \\ \hline
        Type & - & The type indicating whether the galaxy is a central ($=0$), a satellite hosted by a subhalo ($=1$), or a satellite that has lost its subhalo ($=2$) in a FOF group. \\ \hline
        HaloIndex & - & The index of the host subhalo in the subvolume \hbt catalog. \\ \hline
        FirstHaloInFOFgroup & - & The index of the main subhalo in an FOF group  halo catalog. \\ \hline
        CentralGal & - & The index of the central galaxy in the subvolume. \\ \hline
        SnapNum & - & The snapshot number. \\ \hline
        CentralMvir & $10^{10} \mathrm{M_\odot}/h$ & The $M_{\rm 200c}$ of the central subhalo, corresponding to the BoundM200Crit field of the central subhalo in the same FoF group in the \hbt table. \\ \hline
        GasSpin &  Mpc$/h \cdot \rm km/s$ & The spin of the cold gas disk. \\ \hline
        StellarSpin & Mpc$/h \cdot \rm km/s$ & The spin of the stellar disk. \\ \hline
        Pos & Mpc/$h$ & The subhalo position in comoving coordinates.\\ \hline
        Vel & $\rm km/s$ & The physical velocity of the subhalo. \\ \hline
        Len & - & The number of bound particles of the host subhalo. \\ \hline
        Mvir &  $10^{10} \mathrm{M_\odot}/h$ & The $M_{\rm 200c}$ of the subhalo for centrals (Type = 0); the Mbound of the subhalo for satellites (Type $>0$). \\ \hline
        InfallMvir & $10^{10} \mathrm{M_\odot}/h$ & The lastest $M_{\rm 200c}$ of the subhalo when it is/was a central. \\ \hline
        Rvir & Mpc/$h$ & The virial radius of the subhalo the galaxy is/was a center of. \\ \hline
        Vvir & $\rm km/s$ & The virial velocity of the subhalo the galaxy is/was a center of. \\ \hline
        Vmax & $\rm km/s$ & Maximum of the circular velocity. It corresponds to the VmaxPhysical field in \hbt table.
         \\ \hline
        ColdGas & $10^{10} \mathrm{M_\odot}/h$ & The mass in cold gas. \\ \hline
        $\mathrm{H\_mol}$ & $10^{10} \mathrm{M_\odot}/h$ & Total molecular gas mass with a fixed composition: 74\% molecular hydrogen (multiply the total by 0.74 to get the molecular hydrogen mass) and 26\% other components (helium, dust, ionized gas)\\ \hline
        StellarMass & $10^{10} \mathrm{M_\odot}/h$ & The mass in stars.\\ \hline
        BulgeMass & $10^{10} \mathrm{M_\odot}/h$ & The mass in stellar bulge. \\ \hline
        HotGas & $10^{10} \mathrm{M_\odot}/h$ & The mass in hot gas. \\ \hline
        EjectedMass & $10^{10} \mathrm{M_\odot}/h$ & The mass of ejected component. \\ \hline
        BlackHoleMass & $10^{10} \mathrm{M_\odot}/h$ & The mass of the central massive black hole. \\ \hline
        NMetalsColdGas & $10^{10} \mathrm{M_\odot}/h$ & A vector containing the masses of metal elements in cold gas, ordered as: C+Mg, O, Fe, and the sum of the remaining metals. \\ \hline
        NMetalsStellarMass & $10^{10} \mathrm{M_\odot}/h$ & Vector of metal masses in stars, ordered as: C+Mg, O, Fe, and the sum of the remaining metals.  \\ \hline
        NMetalsBulgeMass & $10^{10} \mathrm{M_\odot}/h$ & Vector of metal masses in stellar bulge, ordered as: C+Mg, O, Fe, and the sum of the remaining metals. \\ \hline
        NMetalsHotGas & $10^{10} \mathrm{M_\odot}/h$ & Vector of metal masses in hot gas, ordered as: C+Mg, O, Fe, and the sum of the remaining metals. \\ \hline
        NMetalsEjectedMass & $10^{10} \mathrm{M_\odot}/h$ & Vector of metal masses in ejected component, ordered as: C+Mg, O, Fe, and the sum of the remaining metals. \\ \hline
        Sfr & $\rm {M_\odot}/yr$ & Star formation rate. \\ \hline
        SfrBulge & $\rm {M_\odot}/yr$ & Star formation rate in bulge. \\ \hline
        RPradius & Mpc$/h$ & The radius out of which the cold gas disk is influenced by ram pressure stripping. \\ \hline
        SfrRing & $\rm {M_\odot}/yr$ & Star formation rates in 21 annulii on the cold gas disk. \\ \hline
        Ring & Mpc$/h$ & Radii of the 21 annulii. \\ \hline
        GasRing & $10^{10} \mathrm{M_\odot}/h$ & Masses of cold gas in the 21 annulii. \\ \hline
        FmolRing & - & Fractions of molecular gas in the 21 annulii. \\ \hline
        RPRing & $10^{10} \mathrm{M_\odot}/h$ & Masses of cold gas removed by ram-pressure stripping from the 21 annulli. \\ \hline
        XrayLum & log$_{10} \rm (erg/s)$ & X-Ray luminosity. \\ \hline
        GasDiskRadius & Mpc$/h$ & Three times the scale length of the cold gas disk. \\ \hline
        StellarDiskRadius & Mpc$/h$ & Three times the scale length of the stellar disk. \\ \hline
        BulgeSize & Mpc$/h$ & The half-mass radius of stellar bulge. \\ \hline
        CoolingRadius & Mpc$/h$ & The cooling radius. \\ \hline
        csst\_nuv\_nodust & mag & The intrinsic AB magnitude in CSST NUV band. \\ \hline
        csst\_u\_nodust & mag & The intrinsic AB magnitude in CSST u band.  \\ \hline
        csst\_g\_nodust & mag & The intrinsic AB magnitude in CSST g band.  \\ \hline
        csst\_r\_nodust & mag & The intrinsic AB magnitude in CSST r band.  \\ \hline
        csst\_i\_nodust & mag & The intrinsic AB magnitude in CSST i band.  \\ \hline
        csst\_z\_nodust & mag & The intrinsic AB magnitude in CSST z band.  \\ \hline
        csst\_y\_nodust & mag & The intrinsic AB magnitude in CSST y band.  \\ \hline
        wise\_w1\_nodust & mag & The intrinsic AB magnitude in WISE w1 band.  \\ \hline
        wise\_w2\_nodust & mag & The intrinsic AB magnitude in WISE w2 band.  \\ \hline
        csst\_nuv\_dust & mag & The dust applied AB magnitude in CSST NUV band. \\ \hline
        csst\_u\_dust & mag & The dust applied AB magnitude in CSST u band.  \\ \hline
        csst\_g\_dust & mag & The dust applied AB magnitude in CSST g band.  \\ \hline
        csst\_r\_dust & mag & The dust applied AB magnitude in CSST r band.  \\ \hline
        csst\_i\_dust & mag & The dust applied AB magnitude in CSST i band.  \\ \hline
        csst\_z\_dust & mag & The dust applied AB magnitude in CSST z band.  \\ \hline
        csst\_y\_dust & mag & The dust applied AB magnitude in CSST y band.  \\ \hline
        wise\_w1\_dust & mag & The dust applied AB magnitude in WISE w1 band.  \\ \hline
        wise\_w2\_dust & mag & The dust applied AB magnitude in WISE w2 band.
\label{SAMtableA}
\end{longtable}
\end{center}

\begin{center}
\begin{longtable}{p{0.18\textwidth}p{0.1\textwidth}p{0.7\textwidth}}
\caption{Data fields of the \hbt subhalo catalogs.}\\
\hline\hline
\multicolumn{1}{l}{Field} & \multicolumn{1}{l}{Unit} & \multicolumn{1}{l}{Description}\\

\endfirsthead

\multicolumn{3}{c}{{\tablename\ \thetable{} -- Continued}} \\
\hline\hline
\multicolumn{1}{l}{Field} & \multicolumn{1}{l}{Unit} & \multicolumn{1}{l}{Description} \\ \hline \hline
\endhead

\hline
\multicolumn{3}{r}{{Continued on next page}} \\
\endfoot

\hline\hline
\endlastfoot

\hline\hline
        TrackId & - & The unique ID of the subhalo in this simulation. Persistent over time. \\ \hline
        Nbound & - & The number of dark matter particles bound to the subhalo. \\ \hline
        Mbound & $10^{10} \mathrm{M_\odot}/h$ & The bound mass of the subhalo, equal to Nbound multiplied by the particle mass. \\ \hline
        HostHaloId & - & The index of the host halo in the FoF catalog. \\ \hline
        Rank & - & The rank of the subhalo within its host group, sorted by mass, with rank=0 indicating the most massive subhalo in each group (i.e., the main/central subhalo), and rank$>$0 for satellites. \\ \hline
        Depth & - & The hierarchical level of the subhalo in its host according to the merger tree, with central=0, sub=1, sub-sub=2... \\ \hline
        LastMaxMass & $10^{10} \mathrm{M_\odot}/h$  & The maximum Mbound of the subhalo in current and all preceding snapshots. \\ \hline
        SnapshotIndexOf\ LastMaxMass & - & The snapshot number at which the subhalo attained its maximum Mbound. \\ \hline
        SnapshotIndexOf\ LastIsolation & - & The snapshot number of the most recent time the subhalo is/was a central. \\ \hline
        SnapshotIndexOfBirth & - & The snapshot number at which the subhalo first became resolved. \\ \hline
        SnapshotIndexOfDeath & - & The snapshot number at which the subhalo first became unresolved; this value is only set if the current snapshot $>=$ SnapshotIndexOfDeath, and is $-1$ otherwise. \\ \hline
        SnapshotIndexOfSink & - & The snapshot number of the subhalo when it sinks into another subhalo. \\ \hline
        RmaxComoving & Mpc/$h$ & The comoving radius at which the circular velocity reaches VmaxPhysical, centred on the ComovingMostBoundPosition.\\ \hline
        VmaxPhysical & $\rm km/s$ & The maximum value of the circular velocity of the subhalo, centred on the ComovingMostBoundPosition.  \\ \hline
        LastMaxVmaxPhysical & $\rm km/s$ & The maximum VmaxPhysical of the subhalo in current and preceding snapshots. \\ \hline
        SnapshotIndexOf\ LastMaxVmax & - & The snapshot number of the subhalo at which it has maximum VmaxPhysical.  \\ \hline
        R2SigmaComoving & Mpc/$h$ & The comoving radius within which 95.5\% of the bound particles are contained. \\ \hline
        RHalfComoving & Mpc/$h$ & The half Mbound radius in comoving distance. \\ \hline
        BoundR200CritCo\ moving & Mpc/$h$ & The $R_{\rm 200c}$ of the subhalo calculated using its bound particles. \\ \hline
        BoundM200Crit & $10^{10} \mathrm{M_\odot}/h$ & The $M_{\rm 200c}$ of the subhalo calculated using its bound particles. \\ \hline
        SpecificSelfPotential\ Energy & - & The average specific potential energy per particle, $\phi/m$. The total potential energy of the system is $0.5 \times (\phi/m) \times \mathrm{Mbound}$, where the factor of 0.5 corrects for the double counting of mutual potential energy. \\ \hline
        SpecificSelfKinetic\ Energy & $\rm (km/s)^2$ & The average specific kinetic energy per particle, in the center of mass reference frame of the subhalo. \\ \hline
        SpecificAngularMo\ mentum & Mpc$/h \cdot \rm km/s$ & The specific angular momentum of the subhalo. It is equal to $\langle \vec{r}_{\mathrm{physical}} \times \vec{v}_{\mathrm{physical}} \rangle$ for the particles in the subhalo, calculated in the center of mass reference frame of the subhalo. \\ \hline
        InertialEigenVector & - & The inertial eigenvectors of the InertialTensor matrix of the subhalo, with amplitudes equal to eigenvalues. \\ \hline
        InertialEigenVector\ Weighted & - & Eigenvectors of the InertialTensorWeighted.\\ \hline
        InertialTensor & $(\mathrm{Mpc}/h)^2$ & Flattened representation of the inertia tensor of the subhalo, ($ I_{xx}, I_{xy}, I_{xz}, I_{yy}, I_{yz}, I_{zz}$), with $I_{ij}=\sum_k m_k x_{k,i} x_{k,j} / \sum_k m_k$, where $x_{k,i}$ is the $i$-th component of the coordinate of particle $k$ in the reference frame of the most bound particle, and $m_k$ is the particle mass. \\ \hline
        InertialTensorWeighted & - & The inverse distance-weighted InertialTensor, with $x_k$ replaced by $x_k/r_k$, where $r_k$ is the distance of particle $k$ from the most bound particle.\\ \hline
        ComovingAveragePo\ sition & Mpc/$h$ & The average position of the bound particles in this subhalo. \\ \hline
        PhysicalAverageVe\ locity & km/s & The average velocity of the bound particles in this subhalo. \\ \hline
        ComovingMostBound\ Position & Mpc/$h$ & Position of the most bound particle of this subhalo. \\ \hline
        PhysicalMostBound\ Velocity & km/s & Velocity of the most bound particle of this subhalo. \\ \hline
        MostBoundParticleId & - & ID of the most bound particle of this subhalo. \\ \hline
        SinkTrackId & - & The TrackId of the subhalo that this subhalo has sunk into. -1 if it has not sunk. \\ \hline
        Group\_mass & $10^{10} \mathrm{M_\odot}/h$ & The sum of the masses of all particles in the host FoF halo. \\ \hline
        Group\_M\_200crit & $10^{10} \mathrm{M_\odot}/h$ & The $M_{\rm 200c}$ of the host FoF halo, calculated using all the particles within the corresponding sphere enclosing an average density of 200 times the critical density. \\ \hline
        Group\_R\_200crit & Mpc/$h$ & The $R_{\rm 200c}$ of the host halo. \\ \hline
        Group\_M\_200mean & $10^{10} \mathrm{M_\odot}/h$ & The $M_{\rm 200mean}$ of the host halo, calculated using all the particles within the corresponding sphere enclosing an average density of 200 times the mean density of the universe. \\ \hline
        Group\_R\_200mean & Mpc/$h$ & The $R_{\rm 200mean}$ of the host halo. \\ \hline
        Group\_M\_200tophat & $10^{10} \mathrm{M_\odot}/h$ & The $M_{\rm 200tophat}$ of the host halo, calculated using all the particles within the corresponding sphere enclosing an average density predicted by the tophat spherical collapse model. \\ \hline
        Group\_R\_200tophat & Mpc/$h$ & The $R_{\rm 200tophat}$ of the host halo.
\label{HBTtableA}
\end{longtable}
\end{center}

\begin{center}
\begin{longtable}{p{0.18\textwidth}p{0.1\textwidth}p{0.7\textwidth}}
\caption{Data fields of the light-cone catalogs.}\\
\hline\hline
\multicolumn{1}{l}{Field} & \multicolumn{1}{l}{Unit} & \multicolumn{1}{l}{Description}\\

\endfirsthead

\multicolumn{3}{c}{{\tablename\ \thetable{} -- Continued}} \\
\hline\hline
\multicolumn{1}{l}{Field} & \multicolumn{1}{l}{Unit} & \multicolumn{1}{l}{Description} \\ \hline \hline
\endhead

\hline
\multicolumn{3}{r}{{Continued on next page}} \\
\endfoot

\hline\hline
\endlastfoot

\hline\hline
        trackID & - & The unique ID of the subhalo in the simulation. \\ \hline
        Nbound & - & The number of self-bound particles of the subhalo. \\ \hline
        hosthaloID & - & The index of the host halo in the FoF catalog. \\ \hline
        rank & - & The order of subhalo inside its host group, sorted by the mass.\\ \hline
        posx & Mpc/$h$ & The X-coordinate of the subhalo position in comoving coordinates. (Interpolated) \\ \hline
        posy & Mpc/$h$ & The Y-coordinate of the subhalo position in comoving coordinates. (Interpolated) \\ \hline
        posz & Mpc/$h$ & The Z-coordinate of the subhalo position in comoving coordinates. (Interpolated) \\ \hline
        velx & $\rm km/s$ & The X-component of the subhalo peculiar velocity. (Interpolated) \\ \hline
        vely & $\rm km/s$ & The Y-component of the subhalo peculiar velocity. (Interpolated) \\ \hline
        velz & $\rm km/s$ & The Z-component of the subhalo peculiar velocity. (Interpolated) \\ \hline
        v\_los & $\rm km/s$ & The line-of-sight velocity of the subhalo.\\ \hline
        shMbound & $10^{10} \mathrm{M_\odot}/h$ & The self-bound mass of the subhalo. \\ \hline
        d\_comoving & Mpc/$h$ & The comoving distance of the subhalo. \\ \hline
        RA\_deg & degree & The right ascension of the subhalo in the light-cone, ranging from 0 to 360 degree. \\ \hline
        Dec\_deg & degree & The declination of the subhalo in the light-cone, ranging from -90 to 90 degree. \\ \hline
        vmax & $\rm km/s$ & The maximum circular velocity. It corresponds to the VmaxPhysical field in the \hbt table. \\ \hline
        PeakMass & $10^{10} \mathrm{M_\odot}/h$ & The maximum bound mass. It corresponds to the LastMaxMass field in the \hbt table. \\ \hline
        PeakVmax & $\rm km/s$ & The maximum vmax. It corresponds to LastMaxVmaxPhysical field in the \hbt table. \\ \hline
        shBoundM200Crit & $10^{10} \mathrm{M_\odot}/h$ & The $M_{200c}$ of the subhalo. It corresponds to BoundM200Crit field in \hbt table. \\ \hline
        redshift\_true & - & The TRUE redshift (cosmological redshift) of the subhalo. \\ \hline
        redshift\_obs & - & The observed redshift of the subhalo, converted using redshift\_true and v\_los.\\ \hline
        snapNum & - & The snapshot number of the endpoint of each segment while interpolating. \\ \hline
        delta\_t & - & The scaled evolution time between two snapshots, ranging from 0 to 1. This value can be used to do interpolations for other properties if needed. \\ \hline
        Type & - & The type indicating whether the galaxy is a central ($=0$), a satellite hosted by a subhalo ($=1$), or a satellite that has lost its subhalo ($=2$) in a FOF group. \\ \hline
        CentralMvir & $10^{10} \mathrm{M_\odot}/h$ & The $M_{\rm 200c}$ of the central subhalo. It corresponds to the BoundM200Crit field in \hbt table. \\ \hline
        CentralGal & - & The TrackId of the central galaxy. \\ \hline
        csst\_nuv\_dust & mag & CSST NUV band AB magnitude with dust. (Interpolated) \\ \hline
        csst\_u\_dust & mag & CSST u band AB magnitude with dust. (Interpolated) \\ \hline
        csst\_g\_dust & mag & CSST g band AB magnitude with dust. (Interpolated) \\ \hline
        csst\_r\_dust & mag & CSST r band AB magnitude with dust. (Interpolated) \\ \hline
        csst\_i\_dust & mag & CSST i band AB magnitude with dust. (Interpolated) \\ \hline
        csst\_z\_dust & mag & CSST z band AB magnitude with dust. (Interpolated) \\ \hline
        csst\_y\_dust & mag & CSST y band AB magnitude with dust. (Interpolated) \\ \hline
        wise\_w1\_dust & mag & WISE w1 band AB magnitude with dust. (Interpolated) \\ \hline
        wise\_w2\_dust & mag & WISE w2 band AB magnitude with dust. (Interpolated)
\label{LCtable}
\end{longtable}
\end{center}


\end{document}